\documentclass[aps,twocolumn,showpacs,prr,amsmath,amssymb,floatfix,superscriptaddress]{revtex4-2}
\usepackage{color}
\usepackage[usenames,dvipsnames]{xcolor}
\usepackage[colorlinks=true,citecolor=Cerulean,linkcolor=RubineRed,urlcolor=Cerulean]{hyperref}
\usepackage[normalem]{ulem}
\usepackage{soul,xcolor}
\usepackage{mathptmx} 
\usepackage{dcolumn}
\usepackage{bm}
\usepackage{amsmath}
\usepackage{amssymb}
\usepackage{subfigure}
\usepackage{ulem}
\usepackage{graphicx}
\usepackage{fontenc}
\usepackage{tipa}
\usepackage{url}
\usepackage{booktabs}
\usepackage{multirow}
\newcommand{\dbar}{\text{\textcrd}}
\newcommand{\dd}{\mathrm{d}}

\newcommand{\E}[1]{\langle#1\rangle}

\begin{document}
	
	\title{ Relation between fluctuations and efficiency at maximum power for small heat engines }
	\author{Guo-Hua Xu}
	\email[]{guohuax@zju.edu.cn}
	\affiliation{Department of Physics and Zhejiang Institute of Modern Physics, Zhejiang University, Hangzhou, Zhejiang 310027, China}
	
	\author{Chao Jiang}
	\affiliation{Department of Physics and Zhejiang Institute of Modern Physics, Zhejiang University, Hangzhou, Zhejiang 310027, China}
	
	\author{Yuki Minami}
	\affiliation{Department of Physics and Zhejiang Institute of Modern Physics, Zhejiang University, Hangzhou, Zhejiang 310027, China}
	
	\author{Gentaro Watanabe}
	\email[]{gentaro@zju.edu.cn}
	\affiliation{Department of Physics and Zhejiang Institute of Modern Physics, Zhejiang University, Hangzhou, Zhejiang 310027, China}
	\affiliation{Zhejiang Province Key Laboratory of Quantum Technology and Device, Zhejiang University, Hangzhou, Zhejiang 310027, China}

	\date{\today}
	
	\begin{abstract}
		We study the ratio between the variances of work output and heat input, $\eta^{(2)}$, for a class of four-stroke heat engines which covers various typical cycles. Recent studies on the upper and lower bounds of $\eta^{(2)}$ are based on the quasistatic limit and the linear response regime, respectively. We extend these relations to the finite-time regime within the endoreversible approximation. We consider the ratio $\eta_{\text{MP}}^{(2)}$ at maximum power and find that the square of the Curzon-Ahlborn efficiency, $\eta_{CA}^2$, gives a good estimate of $\eta_{\text{MP}}^{(2)}$ for the class of heat engines considered, i.e., $\eta_{\text{MP}}^{(2)} \simeq \eta_{CA}^2$. This resembles the situation where the Curzon-Ahlborn efficiency gives a good estimate of the efficiency at maximum power for various kinds of finite-time heat engines. Taking an overdamped Brownian particle in a harmonic potential as an example, we can realize such endoreversible small heat engines and give an expression of the cumulants of work output and heat input. The approximate relation $\eta_{\text{MP}}^{(2)} \simeq \eta_{CA}^2$ is verified by numerical simulations. This relation also suggests a trade-off between the efficiency and the stability of finite-time heat engines at maximum power.
	\end{abstract}
	
	\maketitle
	
	\section{introduction}
Relation between the work output and heat input of heat engines is an important issue in thermodynamics. If the heat engine converts the heat input $Q_h$ from a hot heat bath to the work output $W$ and emits heat $Q_c$ to a cold heat bath, the performance of the engine is characterized by the efficiency,
	\begin{equation}
		\eta \equiv \frac{\E{W}}{\E{Q_h}} .
	\end{equation}
Here, $\langle \cdots \rangle$ is the ensemble average. Suppose the temperatures of the hot and cold baths are $T_h$ and $T_c$, respectively, $\eta$ is upper bounded by the Carnot efficiency $\eta_C$ given by $\eta_C \equiv 1- (T_c/T_h)$.
	
With the development of technology in recent decades, small heat engines can be realized in the sub-micron scales \cite{Hugel,Steeneken,Shoichi,Blickle_experiment,Brownian_Carnot,Krishnamurthy,Argun,Colloidal_heat_engines,Exp_review,Exp_molecular}. 
Among them, a typical example is the experimental realization of the so-called Brownian heat engine \cite{Blickle_experiment,Brownian_Carnot,Krishnamurthy,Argun} which consists of a Brownian particle subject to a time-dependent laser trap. 
For the small thermal devices, thermodynamic quantities like work, heat, and entropy production are random variables defined on individual trajectories in the phase space of working substance \cite{sekimoto_book,Seifert_2012,Seifert_2019,Jarzynski_review}. Due to their small number of degrees of freedom, such small thermal devices have large fluctuations of thermodynamic quantities \cite{Noneq_smallsys,Ciliberto_fluct}. These fluctuations can significantly affect the performance of small heat engines. Thus, it is important to study the fluctuations of work output and heat input, which reflect the stability of small heat engines.

Concerning fluctuations of the thermodynamic quantities, the performance of small heat engines has been studied \cite{Sinitsyn_2011,Lahiri_2012,Campisi_2014,Single-particle_HE,Otto_powerlaw,ito2019universal,Saryal_fluct_qOtto,saryal2021universal,Miller_geo,watanabe2021finitetime,Holubec_2021,chen2021microscopic,work_fluct,Work_distribution}. Several features of the fluctuations have been revealed, such as the statistics of stochastic efficiency $\tilde{\eta} \equiv W/Q_h$ \cite{stocastic_efficiency,stocastic_efficiency_2,Stoch_eta_3,SE_4,SE_5,SE_6,SE_7} and the thermodynamics uncertainty relation (TUR)  \cite{original_TUR,proof_ori_TUR,review_TUR,Pietzonka_Trade-Off,Holubec_CTUR,modified_EP_1,modified_EP_2,Operationally_Accessible} which is the relation between the uncertainty (relative fluctuation) of thermodynamic quantities and the entropy production. More precisely, the TUR gives a lower bound of the uncertainty $\sigma_{X} = (\langle X^2 \rangle - \langle X \rangle^2)/\langle X \rangle^2$ for a thermodynamic quantity $X$. 
Recently, a universal bound has been found for microscopic heat engines in the quasistatic limit \cite{ito2019universal}:
 \begin{equation}
 	\eta^{(2)} \equiv \frac{\E{(\Delta W)^2}}{\E{(\Delta Q_h )^2}} \le \eta_C^2 \label{Ito}
 \end{equation}
with $\Delta X \equiv X - \langle X \rangle$ for $X=W$ and $Q_h$.  
The ratio $\eta^{(2)}$ is the relative fluctuation between the work output and the heat input. It is noted that the smaller $\eta^{(2)}$ gives more stable work output converting from the fluctuating heat input. Thus, $\eta^{(2)}$ reflects the stability of small heat engines.

Furthermore, recent works on $\eta^{(2)}$ also suggested its lower bound, $\eta^{(2)} \ge \eta^2$, in the linear response regime of small change in parameters \cite{saryal2021universal,Saryal_continuous} or for a special model of the finite-time quantum Otto engine \cite{Saryal_fluct_qOtto}. However, the relation between $\eta^{(2)}$ and $\eta^2$ for general cycles is still unclear even in the quasistatic limit. In addition, another way to interpret this lower bound is $\sigma_{W} \ge \sigma_{Q_h}$ which is complementary to the TURs. Therefore, the relation between $\eta^{(2)}$ and $\eta^2$ is an interesting issue for the statistics of the engine performance. 

Our goal in the present work is twofold. The first one is to find a relation between $\eta^{(2)}$ and $\eta^2$ in the quasistatic limit, and the second one is to evaluate $\eta^{(2)}$ and $\eta$ for the finite cycle period. 
For practical heat engines with nonzero power output, the discussion in the quasistatic limit is not enough. In the non-quasistatic regime, endoreversible thermodynamics made useful assumptions originally for the macroscopic irreversible heat engines (see, e.g., Refs.~\cite{novikov1958efficiency,curzon1975efficiency,Hoffman_review,Andresen_review}). In this case, the working substance is assumed to be reversible and the irreversibility is caused solely by the irreversible heat flow between the heat bath and the working substance. Recently, the endoreversible assumptions are generalized for the microscopic heat engines \cite{chen2021microscopic,bouton2021quantum}. For example, authors of Ref.~\cite{chen2021microscopic} discussed the microscopic endoreversible Curzon-Ahlborn (CA) heat engine whose working substance is a highly underdamped Brownian particle. In this case, the irreversible heat exchange is caused by the interaction between the working substance and the heat bath with different temperatures.

In the framework of endoreversible thermodynamics, finite-time heat engines are usually characterized by the performance at maximum power. For example, the efficiency at maximum power (EMP), $\eta_{\text{MP}}$, is an important quantity which has been studied for various kinds of small heat engines \cite{Schmiedl_2007,DechantPRL2015,DechantEPL2017,Deffner_entropy,ChenJF_2,ChenJF_1,Izumida_2204.00807}. Besides, the Curzon-Ahlborn efficiency, $\eta_{CA}$, gives a good estimate of $\eta_{\text{MP}}$. It is natural to ask what the fluctuation of performance at maximum power is. To answer this question, we study $\eta_{\text{MP}}^{(2)}$ at maximum power. 

This paper is organized as follows. 
In Sec.~\ref{sec_2}, we give the setup of four-stroke heat engines consisting of two adiabatic strokes and two heat transfer strokes. This setup includes various types of cycles, e.g., the Otto, Brayton, and Diesel cycles \cite{Landsberg}.
In Sec.~\ref{sec_3}, we consider reversible small heat engines and assume that the heat capacities are constant in the heat transfer strokes. We discuss the relation between $\eta^{(2)}$ and $\eta^2$, and find a lower bound of the uncertainties of work and heat. 
In Sec.~\ref{sec_4}, based on the endoreversible assumption, an approximate relation, $\eta_{\text{MP}}^{(2)} \simeq \eta_{CA}^2$, is proposed for endoreversible small heat engines. 
In Sec.~\ref{sec_5}, we discuss the endoreversibility of the Brownian heat engine which consists of an overdamped Brownian particle in a harmonic potential. 
In Sec.~\ref{sec_6}, using the setup introduced in Sec.~\ref{sec_5}, we show two examples of the endoreversible small heat engines, the Brownian Otto engine and Brownian CA engine. Our main result, $\eta_{\text{MP}}^{(2)} \simeq \eta_{CA}^2$, is examined numerically.
The structure of this paper and where each result is presented are summarized in Table~\ref{content}.

\begin{table}[t]
	\centering
	\caption{Main contents of this paper.}
	\label{content}
	\begin{tabular}{cccc}
		\toprule
& section &	           results         & Eqs. \& Fig. \\
		\midrule
\multirow{2}{*}{reversible} 

& \ref{sec_3a} &	  relation between $\eta^{(2)}$ and $\eta^2$ & \eqref{eta_rev} \eqref{eta2 etaC} \eqref{eta2_rev_gen} \\ 

& \ref{sec_3b} &   lower bound of uncertainties              & \eqref{lowerbound}  \\
\midrule
\multirow{2}{*}{endoreversible} 

& \ref{sec_4a} &   $\eta_{\text{MP}}$  at maximum power           & \eqref{EMP_fixalpha}\\

& \ref{sec_4b} &	  $\eta^{(2)}_{\text{MP}}$ at maximum power          & \eqref{approx_1} \\
\midrule
\multirow{2}{*}{Brownian} 

& \ref{sec_5a} &   cumulant of work and heat                 & \eqref{cumulant} \\
	
& \ref{sec_5b} &	  endoreversibility                                 & \eqref{endo_Brownian} \\

\cline{2-4}
\multirow{3}{*}{Otto}
& \ref{sec_6a1} & $\eta_{\text{MP}}$           & \eqref{eta_MP_CA_Otto}      \\
& \ref{sec_6a2} & $\eta^{(2)}_{\text{MP}}$  &  \eqref{eta2_Otto} \\
& \ref{sec_6a3} & cumulant of work             & \eqref{cumulant_otto}  \\

\cline{2-4}
\multirow{2}{*}{Curzon-Ahlborn}
& \ref{sec_6b1} & $\eta_{\text{MP}}$           & \eqref{EMP_CA} \\
& \ref{sec_6b2} & $\eta^{(2)}_{\text{MP}}$  & Fig. \ref{fig:eta2_CA} \\

		\bottomrule
	\end{tabular}
\end{table}

	\section{setup} \label{sec_2}
\begin{figure}[t]
	\includegraphics[width=1 \columnwidth]{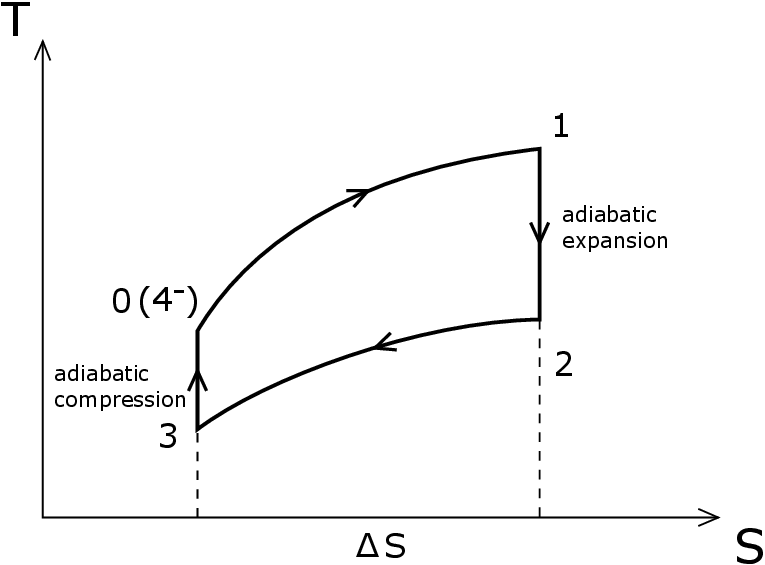} 
	\caption{$T$-$S$ diagram of the class of four-stroke cycles considered in the present work. The cycles consist of two adiabatic strokes (adiabatic expansion $1 \to 2$ and adiabatic compression $3 \to 4^-$) and two heat transfer strokes ($0 \to 1$ and $2 \to 3$). }
	\label{fig:TS}
\end{figure}

In this paper, we consider a small heat engine operating with two heat baths with the temperatures $T_h$ and $T_c$ ($T_h>T_c$), and focus on a class of four-stroke cycles consisting of two adiabatic strokes ($1 \to 2$ and $3 \to 4^-$) and two heat transfer strokes ($0 \to 1$ and $2 \to 3$). The cycle on the $T$-$S$ plane is shown in Fig.~\ref{fig:TS}. Here, $T$ and $S$ are the temperature and the entropy of the working substance, respectively. The final node $4^-$ of the cycle is statistically equivalent to the initial node $0$, i.e., the phase space probability density function (PDF) of the working substance at these nodes are the same. During the heat transfer stroke $0 \to 1$ ($2 \to 3$), heat flows into (out of) the engine from the hot heat bath (to the cold bath). Throughout the paper, we take the sign convention where $Q_h$ is positive when the heat is absorbed by the engine from the hot bath and $Q_c$ is positive when the heat is emitted from the engine to the cold bath, i.e., $\E{Q_i} > 0$ ($i=h$ and $c$). 
	
Our goal is to derive the relation between the ratio $\eta^{(2)} \equiv {\E{(\Delta W)^2}}/{\E{(\Delta Q_h )^2}}$ and the efficiency $\eta \equiv \E{W} / \E{Q_h}$ for various kinds of cycles not only in the quasistatic regime but also in the finite-time regime.
	
	\section{  Reversible small heat engine } \label{sec_3}

In this section, we compare $\eta^{(2)}$ and $\eta$ and study the uncertainties of work and heat in the reversible case \footnote{The term ``reversible'' in this work means that the working substance of the engine is reversible such that the working substance is always in a canonical state for some temperature. Following the assumption of endoreversible thermodynamics, we regard that the working substance is reversible, but the heat flow between the working substance and the bath can be irreversible.}. We consider the working substance following a reversible cycle. The parameter change of the working substance is much slower than the relaxation, so that the working substance is always in the (local) equilibrium state during the heat transfer strokes. 

We make the following three assumptions on the four-stroke small heat engines: 

(i) We consider particular types of the heat transfer stroke where the amount of the heat exchange between the engine and the bath can be described by the heat capacity $C_i$ and the temperature change $\dd T$ of the working substance in the form of
\begin{equation}
	\langle \dbar Q_i \rangle=| C_i \dd T | . \label{capacity}
\end{equation}
Here, $i = h~(c)$ for the heat transfer stroke $0\to 1$ ($2 \to 3$) with the hot (cold) heat bath. Such heat transfer strokes include isochoric strokes and isobaric strokes. Thus, various typical cycles such as the Otto, Diesel, and Brayton cycles are covered in the present discussion. For example, the Otto engine absorbs (emits) the heat during the hot (cold) isochoric stroke with the heat capacity $C_h$ ($C_c$) at constant volume, and the Diesel engine absorbs the heat during the isobaric stroke with the heat capacity $C_h$ at constant pressure and emits the heat during the isochoric stroke with the heat capacity $C_c$ at constant volume. 

(ii) We further assume that the heat capacity $C_i$ of the working substance is constant (and positive) during each heat transfer stroke \footnote{There are many cases that the heat capacities of the heat transfer strokes are constant. For example, isobaric and isochoric strokes for a Brownian particle in a harmonic oscillator potential or a square-well potential. }. In addition to $C_i$, we assume that the heat capacity $C_V$ at constant volume, $C_V \equiv (\partial \E{E}/\partial T)_V$, is also constant during each heat transfer stroke and does not depend on the volume. Since the heat capacity $C_i$ is positive, we always have $T_1-T_0>0$ and $T_3-T_2<0$. To avoid the situation in which the two heat transfer strokes cross on the $T$-$S$ plane, we consider the cases with $T_1\ge T_2$ and $T_0\ge T_3$. If the two heat transfer strokes cross each other, the whole cycle can be decomposed into two sub-cycles with the clock-wise and counter-clock-wise directions on the $T$-$S$ plane corresponding to a heat engine and a refrigerator, respectively. 

(iii) We consider reversible cycles which consist of reversible heat transfer strokes and quasistatic adiabatic strokes without irreversibility at the connections between them. 
In the present case, the quasistatic adiabatic strokes start from a canonical state at the temperature $T_\text{init}$ because the working substance is always in the equilibrium state during the preceding reversible heat transfer stroke. 
Since the number of degrees of freedom in the working substance of microscopic heat engines is small, the energy distribution of the working substance is generally not a canonical distribution at the end of the quasistatic adiabatic strokes.
When the engine is coupled to a heat bath at the temperature $T_\text{fin}$ after an adiabatic stroke, irreversible heat exchange occurs unless the working substance is already in the canonical state at $T_\text{fin}$. A necessary and sufficient condition for the initial state and the final state of the quasistatic adiabatic stroke to be canonical states at the temperature $T_\text{init}$ and $T_\text{fin}$, respectively, is
	\begin{equation}
		\frac{E_\text{init}}{T_\text{init}}=\dfrac{E_\text{fin}}{T_\text{fin}},	 \label{adiabatic_rev}
	\end{equation}
up to a constant \cite{sato_rev_cond}. Here, $E_\text{init}$ ($E_\text{fin}$) is the internal energy of the initial (final) state of the quasistatic adiabatic stroke. Note that $E_\text{init}$ and $E_\text{fin}$ are random variables and Eq.~\eqref{adiabatic_rev} has to hold for each realization.

\subsection{Relation between $\eta^{(2)}$ and $\eta$} \label{sec_3a}

Based on the above mentioned assumptions, we can derive the variances of work output and heat input. Work output during the quasistatic adiabatic strokes $1 \to 2$ and $3\to 4^-$ is given by
	\begin{equation}
		W_{1\to 2}=E_1-E_2 = \bigg(1- \dfrac{T_2}{T_1} \bigg) E_1 
	\end{equation}
and
	\begin{equation}
		W_{3\to 4^-}=E_3-E_{4^-} = \bigg(\dfrac{T_3}{T_0} -1 \bigg) E_{4^-} ,
	\end{equation}
respectively, with $E_j$ and $T_j$ being the internal energy and the temperature of the working substance at node $j$, respectively. Here, we have assumed the  reversibility condition \eqref{adiabatic_rev} for the adiabatic strokes $1 \to 2$ and $3 \to 4^-$:
	\begin{equation}
		\frac{E_1}{T_1}=\dfrac{E_2}{T_2} ~\text{~~~~and~~~~}~ \frac{E_3}{T_3}=\dfrac{E_{4^-}}{T_0}.
	\end{equation}
In the quasistatic limit, fluctuations of work during the heat transfer strokes $0\to 1$ and $2\to 3$ are negligible due to the same reason for quasistatic isothermal processes \cite{sekimoto_book,ito2019universal,Holubec_CTUR}:
	\begin{equation}
		\Delta W_{0 \to 1} \equiv W_{0 \to 1} - \E{W_{0 \to 1} } \simeq 0
	\end{equation}
and 
\begin{equation}
	\Delta W_{2 \to 3} \equiv W_{2 \to 3} - \E{W_{2 \to 3} } \simeq 0.
\end{equation}
In addition, since the quasistatic adiabatic strokes $1\to 2$ and $3 \to 4^-$ are separated by a reversible heat transfer stroke, their work output $W_{1\to 2}$ and $W_{3\to 4^-}$ are not correlated. Therefore, the variance of work output in one cycle is given by the sum of the variances in the uncorrelated quasistatic adiabatic strokes:
	\begin{align}
		\langle (\Delta W)^2 \rangle 
		&= \langle (\Delta W_{1\to 2})^2 \rangle + \langle (\Delta W_{3\to 4^-})^2 \rangle \notag \\ 
		& = \bigg(1- \frac{T_2}{T_1} \bigg)^2 \langle (\Delta E_1)^2 \rangle + \bigg( 1- \frac{T_3}{T_0}\bigg)^2 \langle (\Delta E_0)^2 \rangle \notag \\
		& = k_B C_V \big[ (T_1-T_2)^2 +(T_0-T_3)^2 \big]. \label{varW}
	\end{align} 
	Here, we have used the property: $\langle (\Delta E_j)^2 \rangle = k_B C_V T_j^2$ \cite{Qian}.
	
	Heat absorbed from the hot bath during the heat transfer stroke $0 \to 1$ reads 
	\begin{equation}
		Q_h = E_1-E_0-W_{0\to 1},
	\end{equation}
and its variance is given by
	\begin{align}
		\langle (\Delta Q_h)^2 \rangle &= \langle (\Delta E_0)^2 \rangle + \langle (\Delta E_1)^2 \rangle \\
		&= k_B C_V \big( T_1^2 +T_0^2 \big) \label{varQh}
	\end{align}
because $E_0$ and $E_1$ are uncorrelated and $\Delta W_{0 \to 1} \simeq 0$. Thus, from Eqs.~\eqref{varW} and \eqref{varQh}, the ratio $\eta^{(2)}$ can be written as
	\begin{equation}
		\eta^{(2)} = \frac{\langle (\Delta W)^2 \rangle }{ \langle (\Delta Q_h)^2 \rangle} = \frac{(T_1-T_2)^2+(T_0-T_3)^2}{T_1^2+T_0^2} \label{eta2 qs}.
	\end{equation} 

For the reversible cycle, the temperatures $\{T_j\}$ of the working substance are not independent. The mean value of the entropy change of the working substance during the heat absorption or the heat emission stroke is given by
	\begin{equation}
		\Delta S_i = (-1)^{j/2} \int_j^{j+1}\frac{\langle \dbar Q_i \rangle}{T}  = C_{i}\ln \frac{T_{j+1}}{T_j},
	\end{equation}
	where $j=0$ for $i=h$ and $j=2$ for $i=c$. Since the cycle is closed, the mean value of the net entropy change over the cycle is zero:
	\begin{equation}
		\Delta S_h + \Delta S_c = C_{h}\ln \frac{T_1}{T_0} + C_{c}\ln \frac{T_3}{T_2} = 0 .
	\end{equation}
Thus, we obtain
\begin{equation}
	\tilde{t}^{\theta} = T_3/T_2 , \label{temp ratio2}
\end{equation}
where $\tilde{t} \equiv T_0/T_1$ is the temperature ratio with $0<\tilde{t}<1$ and $\theta \equiv C_{h} / C_{c}$ is the heat capacity ratio with $\theta>0$. 

The working substance is in local equilibrium with temperature $T_c<T<T_h$, as assumed in endoreversible thermodynamics. In the whole cycle, the maximum (minimum) temperature of the working substance is $T_1$ ($T_3$). Since the durations of the heat transfer strokes are sufficiently long, we have $T_1=T_h$ and $T_3=T_c$. Then, the Carnot efficiency is given by
\begin{equation}
	\eta_C = 1 - \frac{T_3}{T_1} . \label{etaC_rev}
\end{equation} 

Now, we can express $\eta^{(2)}$ and $\eta$ with three parameters: $\theta$, $\tilde{t}$, and $\eta_C$.
The efficiency is given by
\begin{equation}
	\eta =1-\frac{\langle Q_{c} \rangle}{\langle Q_{h} \rangle}=1-\frac{T_2-T_3}{\theta(T_1-T_0)}   \label{eta qs}.
\end{equation}
Substituting Eqs.~\eqref{temp ratio2} and \eqref{etaC_rev} into Eq.~\eqref{eta qs}, we get 
\begin{equation}
	\eta = 1 - \frac{(1-\eta_C)(1-\tilde{t}^{\theta})}{ \theta(1-\tilde{t})\tilde{t}^{\theta} } \label{eta_rev} ,
\end{equation}
and substituting Eqs.~\eqref{temp ratio2} and \eqref{etaC_rev} into Eqs.~\eqref{eta2 qs}, we get
\begin{equation}
	\eta^{(2)} = \bigg( 1-\frac{1-\eta_C}{\tilde{t}^{\theta}} \bigg)^2\frac{1}{1+\tilde{t}^2} + \bigg( 1-\frac{1-\eta_C}{\tilde{t}} \bigg)^2\frac{\tilde{t}^2}{1+\tilde{t}^2} .  \label{eta2 etaC} 
\end{equation}
Since $0<\tilde{t}<1$ and $\theta>0$, $\eta^{(2)}$ is upper bounded by $\eta_C^2$:
\begin{equation}
	\eta^{(2)}<\eta_C^2,
\end{equation}
which is consistent with the universal bound on $\eta^{(2)}$ proven for general quasistatic cycles in our previous work \cite{ito2019universal}. It is interesting to note that, although the present discussion leading to Eq.~\eqref{eta2 etaC} does not include the Carnot cycle, the resulting Eq.~\eqref{eta2 etaC} for $\tilde{t}=1$ and $\theta=1$ is consistent with $\eta^{(2)}$ of the Carnot cycle, $\eta^{(2)} = \eta_C^2$, obtained in Ref.~\cite{ito2019universal}. 

\begin{figure}[t]
	\includegraphics[width=1 \columnwidth]{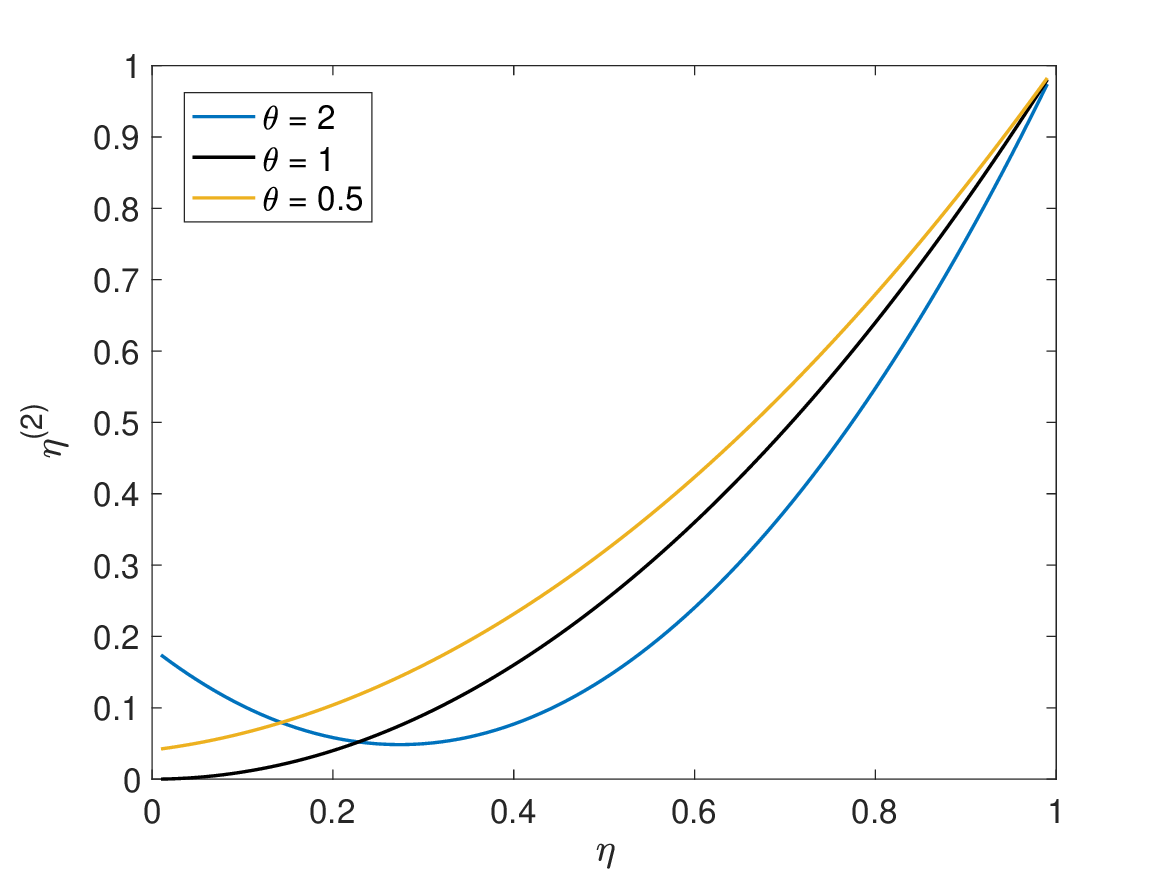} 
	\caption{$\eta^{(2)}$ as a function of $\eta$ for $\theta = 0.5$, $1$, and $2$. Here, we set $\tilde{t} = 0.4$. The black line ($\eta^{(2)}$ for $\theta=1$) shows $\eta^{(2)}=\eta^2$. }
	\label{fig:eta2_eta_qs}
\end{figure}

Next, we compare the values of $\eta^{(2)}$ and $\eta^2$. 
For $\theta=1$, which means that the two heat transfer strokes in the cycle are the same type (e.g., the Otto cycle and the Brayton cycle), we readily get 
\begin{equation}
	\eta^{(2)}=\eta^2 \label{eta2_rev_gen}
\end{equation}
from Eqs.~\eqref{eta_rev} and \eqref{eta2 etaC} with
\begin{equation}
	\eta = 1 - \frac{T_3}{T_0} = 1 - \frac{T_2}{T_1} . \label{eta_theta1}
\end{equation}
For $\theta \neq 1$, as shown in Fig.~\ref{fig:eta2_eta_qs}, $\eta^{(2)}$ is a shifted quadratic function of $\eta$.
For $0<\theta<1$, we have 
\begin{equation}
	\eta^{(2)}  > \eta^2 ,
\end{equation}
which is consistent with the lower bound in the linear response regime of small change in parameters \cite{saryal2021universal,Saryal_continuous}. However, for $\theta>1$, $\eta^2$ does not give the lower bound of $\eta^{(2)}$ any more, and $\eta^{(2)}$ could be larger or smaller than $\eta^2$. This result can also be interpreted as the relation between uncertainties $\sigma_{W}$ and $\sigma_{Q_h}$ of the work output and the heat input from the hot heat bath, which are defined as $\sigma_{X} = (\langle X^2 \rangle - \langle X \rangle^2)/\langle X \rangle^2$ with $X=W$ and $Q_h$. We have
\begin{equation}
	\sigma_W = \sigma_{Q_h} 
\end{equation}
when $\theta = 1$, and 
\begin{equation}
	\sigma_W > \sigma_{Q_h} \label{ineqWQ}
\end{equation}
when $0<\theta<1$. $\sigma_{W}$ could be larger or smaller than $\sigma_{Q_h}$ when $\theta>1$.

\subsection{Uncertainties of work output and heat input} \label{sec_3b}

Now, we evaluate the uncertainties of work output and heat input and give a lower bound of them. 
Before discussing the four-stroke cycles with constant $C_i$, let us first consider the Carnot cycle as a simple example which gives $\eta^{(2)}=\eta_C^2$ and $\sigma_W = \sigma_{Q_h}$ \cite{ito2019universal}. In this case, we have $\langle (\Delta Q_h)^2 \rangle = 2k_B C_V T_h^2$ and $\langle Q_h \rangle= T_h \Delta S$ with $\Delta S \equiv \Delta S_h = -\Delta S_c > 0$, and the uncertainty $\sigma_{Q_h} $ is given by
	\begin{equation}
		\sigma_{Q_h} = \frac{\langle (\Delta Q_h)^2 \rangle}{\langle Q_h \rangle^2} = \frac{2 k_B C_V}{(\Delta S)^2} . \label{work_fluct}
	\end{equation}
The Carnot engine whose working substance is an overdamped Brownian particle trapped in the potential $U = k \, x^{2n}/2n$ ($n$ is a natural number) discussed in Ref.~\cite{Holubec_CTUR} is a special case of Eq.~\eqref{work_fluct} \footnote{The Hamiltonian in Ref.~\cite{Holubec_CTUR} is $H=kx^{2n}/2n$, where $x$ is the position of the Brownian particle, $k$ is the stiffness of the potential, and $n$ is a natural number. In this case, $C_V$ is constant, $C_V = k_B/(2n)$, which is in accordance with the assumption of our analysis. Therefore, Eq.~\eqref{work_fluct} is applicable to this case, and gives  $\sqrt{\sigma_W} = k_B/(\sqrt{n}\,\Delta S)$, which agrees with the result in Ref.~\cite{Holubec_CTUR}. }. 

Next, let us go back to the reversible four-stroke cycles with constant $C_i$ in the heat transfer strokes $0\to1$ and $2\to3$. According to Eqs.~\eqref{capacity} and \eqref{varQh}, the mean value of the absorbed heat from the hot heat bath is given by $\langle Q_h \rangle = C_{h}(T_1-T_0)$, and its variance is given by $\langle (\Delta Q_h)^2 \rangle = k_B C_V ( T_1^2 +T_0^2 )$. Thus, we get 
\begin{equation}
	\sigma_{Q_h} 
	= \frac{k_B C_V}{C_h^2} \frac{1 + \tilde{t}^2}{( 1 - \tilde{t})^2}, \label{uncertainty_Q_h}
\end{equation}
and see that $\sigma_{Q_h}$ is a function of the temperature ratio $\tilde{t} = T_0/T_1$ which is related to the change $\Delta S$ of the entropy: 
\begin{equation}
	\tilde{t} =\exp(-\Delta S/C_h). \label{temp_ratio_10}
\end{equation} 
In the same way, we have
\begin{equation}
	\sigma_{Q_c} = (k_B C_V/C_c^2)(T_2^2+T_3^2)(T_2-T_3)^{-2} \label{uncertainty_Q_c}
\end{equation}
which depends on the temperature ratio $T_2/T_3$ with
\begin{equation}
	T_2/T_3=\exp(\Delta S/C_c). \label{temp_ratio_23}
\end{equation}
From Eqs.~\eqref{uncertainty_Q_h}, \eqref{temp_ratio_10}, \eqref{uncertainty_Q_c}, and \eqref{temp_ratio_23}, we obtain the uncertainty of the absorbed/emitted heat $\sigma_{Q_i}$ ($i = h,c$) and find its lower bound given by 
\begin{equation}
	\sigma_{Q_i} = \dfrac{k_B C_V}{2C_i^2} \dfrac{\cosh(\Delta S/C_i)}{\sinh^2[\Delta S/(2C_i)]} \geq \frac{2 k_B C_V}{(\Delta S)^2}.
\end{equation}
For $0<\theta\le 1$, from Eq.~\eqref{ineqWQ}, we have $\sigma_W\ge\sigma_{Q_h}\ge 2 k_B C_V/(\Delta S)^2$, where the first equality is satisfied when $\theta = 1$ and the second equality is satisfied when $\Delta S \ll C_h$. For $\theta>1$, $\sigma_{W}$ could be larger or smaller than $\sigma_{Q_h}$ but has the same lower bound.
In conclusion, we have
\begin{equation}
		\sigma_{Q_i} \geq \frac{2 k_B C_V}{(\Delta S)^2} \text{~~and~~} \sigma_{W} \geq \frac{2 k_B C_V}{(\Delta S)^2}. \label{lowerbound}
\end{equation}
This lower bound is given by the uncertainty of work output and absorbed/emitted heat in the corresponding Carnot cycle with the same value of $\Delta S$. 
	
	\section{Endoreversible small heat engine}  \label{sec_4}
	
In this section, we consider the small heat engine operating in a finite cycle period $\tau$ which saitisfies the endoreversible assumptions. To show a subtle difference between the endoreversible assumptions in the microscopic and macroscopic systems, we first introduce endoreversible thermodynamics for macroscopic engines before giving our main result for the endoreversible small heat engines. At the end of this section, we derive $\eta_{\text{MP}} = \eta_{CA}$ [Eq.~\eqref{EMP_fixalpha}] for a particular class of small heat engines, such that the two heat transfer strokes are the same type, i.e., $\theta=1$, with the constant heat capacities and heat conductivities.

	\subsection{ Endoreversible thermodynamics for macroscopic engines} \label{sec_4a}

In endoreversible thermodynamics, we assume that the irreversibility is caused solely by the irreversible heat transfer between the engine and the bath, and the other processes are reversible \cite{Hoffman_review,Andresen_review}. For macroscopic endoreversible engines, we assume that the relaxation time of the working substance is much shorter than the timescale of the parameter change. Thus, the working substance is in local equilibrium with an internal temperature $T_{\text{in}}$. This internal temperature is generally different from the temperature of the baths. In addition, the adiabatic strokes are assumed to be reversible but operated in a short time compared with the cycle period. The working substance follows a reversible cycle, and the mean value of the entropy production $\Sigma$ should be zero:
	\begin{equation}
	\Sigma \equiv \Delta S - \bigg\langle \int_0^{\tau} \frac{\dot{Q}}{T_{\text{in}}} \dd t \bigg\rangle = 0 . \label{EP}
\end{equation} 
Here, $\Delta S$ is the change of the entropy in the working substance through one cycle, which is zero for the engine running periodically. For small heat engines, we consider the same assumption of endoreversibility but from a different microscopic mechanism.

For macroscopic engines, there is usually an intermediate heat conducting medium between the working substance and the bath during the heat transfer strokes. Due to the finite heat flow and the finite duration of the heat transfer strokes, there is a temperature difference between the bath and the working substance. The mean value of the heat flow between the bath and the working substance is assumed to be proportional to the temperature difference between them (i.e., the Fourier or the Newton law). Since we consider sufficiently large timescale such that the heat flow has already reached a steady state, the average heat flux from the bath to the heat conducting medium and that from the heat conducting medium to the working substance are in balance. Therefore, within this formalism, we ignore the presence of nonzero mean value of the energy stored in the heat conducting medium. Furthermore, in the standard setup of the small heat engines, there is no intermediate heat conducting medium as we will discuss later.

If the working substance follows the Carnot cycle and the heat transfer follows the Newton law, the engine is referred to as the Curzon-Ahlborn heat engine \cite{curzon1975efficiency}. The internal temperatures $T_{ih}$ and $T_{ic}$ during the hot and cold isothermal strokes give the efficiency of the Carnot engine as $\eta = 1 - (T_{ic}/T_{ih})$. According to the Newton law, the mean value of heat absorbed from the hot heat bath is given by 
\begin{equation}
	\E{Q_h} = \alpha_h \tau_h (T_h-T_{ih})
\end{equation}
with the conductivity $\alpha_h$ and the duration $\tau_h$ of the hot isothermal stroke, and the mean value of heat emitted to the cold heat bath with the duration $\tau_c$ and the conductivity $\alpha_c$ is given by $\E{Q_c} = \alpha_c \tau_c (T_{ic}-T_c)$. To maximize the power under given conductivities ($\alpha_h$ and $\alpha_c$) and bath temperatures ($T_h$ and $T_c$), we optimize the internal temperatures ($T_{ih}$ and $T_{ic}$) and the durations ($\tau_h$ and $\tau_c$). Then, we obtain the optimum condition for $T_{ih}$ and $T_{ic}$ (see, e.g., Ref.~\cite{Hoffman_review} for detailed derivation): 
\begin{equation}
	\frac{T_{ic}^*}{T_{ih}^*} = \sqrt{\frac{T_c}{T_h}} ,
\end{equation}
where $T_{ih}^*$ and $T_{ic}^*$ are the optimum values of $T_{ih}$ and $T_{ic}$. Therefore, the efficiency at maximum power $\eta_{CA}$ of the CA engine is given by \cite{Hoffman_review}
\begin{equation}
	\eta_{CA} = 1-\sqrt{\frac{T_c}{T_h}} .
\end{equation}

In the present paper, we consider not only the CA engines but also the class of four-stroke heat engines introduced in the previous section. In the latter case, during each heat transfer stroke, the internal temperature changes while the bath temperatures $T_h$ and $T_c$ are constant, and the heat transfer follows the Fourier law. Following the derivation in Ref.~\cite{Deffner_entropy}, we find that, if the two heat transfer strokes in the cycle are the same type ($\theta=1$, e.g., the Otto cycle and the Brayton cycle), the efficiency $\eta_{\text{MP}}$ at maximum power is given by
\begin{equation}
	\eta_{\text{MP}}=\eta_{CA}. \label{EMP_fixalpha}
\end{equation}
We also obtain this result for the endoreversible small heat engines with constant conductivities $\alpha_c$ and $\alpha_h$. The detailed derivation is shown at the end of this section. 

	\subsection{Endoreversible thermodynamics for small heat engines} \label{sec_4b}
	
Now, we consider endoreversible small heat engines. We assume a special class of working substance such that its energy distribution is always given by the canonical distribution function with the effective temperature $\tilde{T}$ even if the working substance is far from equilibrium with the heat bath in the irreversible heat transfer strokes. It is noted that the effective temperature $\tilde{T}$ is generally different from the bath temperature. We also assume direct heat transfer between the working substance and the heat bath, and there is no heat conducting medium between them. This situation is possible when the particles in the bath directly interact with the particles of the working substance such as Brownian heat engines \cite{Blickle_experiment,Brownian_Carnot,Argun,Schmiedl_2007}  and the interaction between them is sufficiently weak. This situation can also be realized in an endoreversible quantum Otto cycle \cite{bouton2021quantum}. Similarly to macroscopic endoreversible thermodynamics, we assume that the adiabatic strokes are reversible and the irreversibility is caused solely by the heat flow due to the difference between $\tilde{T}$ and the bath temperature. If the mean value of the heat transfer is linear with respect to the temperature difference, one can reproduce the CA efficiency for the engine operating in the Carnot cycle. One example of microscopic CA engines has been discussed in Ref. \cite{chen2021microscopic} which considers a highly underdamped Brownian particle as the working substance. 

In our work, we study the fluctuation ratio $\eta_{\text{MP}}^{(2)}$ at maximum power for endoreversible small heat engines. In the quasistatic limit, as discussed in Sec.~\ref{sec_3}, the fluctuations of work and heat depend solely on the energy fluctuation at each node, and the correlations of work and heat in each stroke are negligible. If the cycle period at maximum power is sufficiently large, the discussion in Sec.~\ref{sec_3} is still applicable. In addition, since there is no heat conducting medium between the working substance and the bath, fluctuation of heat input to the working substance is exactly equal to fluctuation of heat output from the bath. Then, our result Eq.~\eqref{eta2_rev_gen} holds and the ratio $\eta_{\text{MP}}^{(2)}$ at maximum power is given by $\eta_{\text{MP}}^{(2)}=\eta_{\text{MP}}^2$. Substituting Eq.~\eqref{EMP_fixalpha} into Eq.~\eqref{eta2_rev_gen}, we have
\begin{equation}
	\eta^{(2)}_{\text{MP}  } = \eta_{CA}^2 = \bigg(1 - \sqrt{\frac{T_c}{T_h}} \bigg)^2. \label{eta2_MP}
\end{equation}
It is noted that Eq.~\eqref{eta2_MP} is obtained for large cycle period and constant heat conductivities. For more general cases, our result Eq.~\eqref{eta2_MP} may not be valid. However, as verified for a microscopic model in the next section, the following approximate relation still holds for various kinds of endoreversible small heat engines:
\begin{equation}
	\eta^{(2)}_{\text{MP}  } \simeq \eta_{CA}^2 , \label{approx_1}
\end{equation}
i.e., $\eta_{CA}^2$ gives a good estimate of $\eta_{\text{MP}}^{(2)}$. 
This resembles the situation where the CA efficiency gives a good estimate of the efficiency at maximum power. Although $\eta_{\text{MP}}$ is not always equal to $\eta_{CA}$, the approximate relation 
\begin{equation}
	\eta_{\text{MP}} \simeq \eta_{CA}  \label{approx_2}
\end{equation}
holds for various kinds of heat engines with reasonable choices of optimization conditions. Similarly, we find that $\eta^{(2)}_{\text{MP}  } \simeq \eta_{CA}^2$ holds for a wide class of finite-time heat engines with large cycle period. In the next section, we will show that the approximate relation $\eta^{(2)}_{\text{MP}  } \simeq \eta_{CA}^2$ is applicable to the endoreversible small heat engines even when the cycle period is very small compare to the correlation time of work and heat.

	\subsection{Derivation of Eq.~\eqref{EMP_fixalpha}}

Finally, we give a derivation of Eq.~\eqref{EMP_fixalpha} for the four-stroke heat engines with constant heat capacity $C \equiv C_{h} = C_{c}$. In this case, the mean value of work output in one cycle is given by 
\begin{equation}
	\E{W} = \langle Q_h \rangle - \langle Q_c \rangle = C(\tilde{T}_1-\tilde{T}_0-\tilde{T}_2+\tilde{T}_3),
\end{equation}
where $\tilde{T}_j$ is the effective temperature at node $j$, and the cycle period is defined by
\begin{equation}
	\tau = \gamma (\tau_h+\tau_c),
\end{equation}
where $\tau_h$ and $\tau_c$ are the duration of heat transfer strokes and $\gamma$ is the ratio of the cycle period to the duration of the heat transfer strokes. (In the usual analysis of endoreversible thermodynamics, where the duration of the adiabatic strokes is assumed to be much shorter than that of the heat transfer strokes, $\gamma$ is set to unity as we shall do later.)
Then the power $P = \E{W} / \tau$ depends on $\{\tilde{T}_j\}$, $\tau_h$, and $\tau_c$. We maximize the power by optimizing $\{\tilde{T}_j\}$, $\tau_h$, and $\tau_c$ under given $T_h$, $T_c$, $\alpha_h$, and $\alpha_c$.
According to the Fourier law, the mean value of the heat flow $ \dot{Q}_i $ for $i=h$ and $c$ is given by
\begin{equation}
	\E{ \dot{Q}_i } = \alpha_i (T_i - \tilde{T} ) .
\end{equation}
The time evolution of $\tilde{T}$ during each heat transfer stroke follows the differential equation: 
\begin{equation}
	\frac{\dd \tilde{T}(t)}{\dd t} = \alpha_i (\tilde{T}(t)-T_i) /C .
\end{equation}
For the stroke $0\to1$, the solution is 
\begin{equation}
	\tilde{T}_1-T_h=(\tilde{T}_0-T_h)\exp(-\alpha_h\tau_h/C), \label{T01}
\end{equation}
and for the stroke $2\to3$, the solution is 
\begin{equation}
	\tilde{T}_3-T_c=(\tilde{T}_2-T_c)\exp(-\alpha_c\tau_c/C). \label{T23}
\end{equation}
Since the working substance follows the reversible cycle, as Eq.~\eqref{EP}, the relation between $\tilde{T}_j$ and $\eta$ is the same as Eq.~\eqref{eta_theta1} in the quasistatic limit discussed in Sec.~\ref{sec_3}. For $\theta=1$, we have
\begin{equation}
	\eta = 1 - \frac{\tilde{T}_3}{\tilde{T}_0} = 1 - \frac{\tilde{T}_2}{\tilde{T}_1} . \label{eta_Teff_theta1}
\end{equation}
Then, from Eqs.~\eqref{T01}, \eqref{T23}, and \eqref{eta_Teff_theta1}, we have four equations for the four variables $\{\tilde{T}_j\}$. Thus, $\tilde{T}_j$ is a function of the efficiency and the durations $\tau_h$ and $\tau_c$, i.e., 
\begin{equation}
	\tilde{T}_j = \tilde{T}_{j}(\eta,\tau_h,\tau_c;T_h,T_c,\alpha_h,\alpha_c) .
\end{equation}
Therefore, optimizing $\{\tilde{T}_j\}$, $\tau_h$, and $\tau_c$ under the constraints of Eqs.~\eqref{T01}, \eqref{T23}, and \eqref{eta_Teff_theta1} is equivalent to optimizing $\eta$, $\tau_h$, and $\tau_c$ with the same constraints.
The power is given by
\begin{equation}
	P = \frac{\langle W \rangle}{\tau} = \frac{2C\eta[(1-\eta) T_h-T_c]}{\gamma(1-\eta)(\tau_h+\tau_c)}\frac{\sinh\left(\dfrac{\alpha_c\tau_c}{2C}\right)\sinh\left(\dfrac{\alpha_h\tau_h}{2C}\right)}{\sinh\left(\dfrac{\alpha_c\tau_c}{2C}+\dfrac{\alpha_h\tau_h}{2C}\right)} , \label{Power}
\end{equation}
which can be separated into a product of the following two functions:
\begin{equation}
	f_1(\eta;T_h,T_c) = \frac{2C\eta[(1-\eta) T_h-T_c]}{\gamma(1-\eta)}
\end{equation}
and 
\begin{equation}
	f_2(\tau_h,\tau_c;T_h,T_c,\alpha_h,\alpha_c) = \frac{\sinh\left(\dfrac{\alpha_c\tau_c}{2C}\right)\sinh\left(\dfrac{\alpha_h\tau_h}{2C}\right)}{\left(\tau_h+\tau_c\right) \sinh\left(\dfrac{\alpha_c\tau_c}{2C}+\dfrac{\alpha_h\tau_h}{2C}\right)} .
\end{equation}
When the power is maximized, we have 
\begin{equation}
	\bigg( \frac{\partial f_1}{\partial \eta} \bigg)_{\{ T_i \}} = \bigg( \frac{\partial f_2}{\partial \tau_h} \bigg)_{\{ T_i \},\{ \alpha_i \} } = \bigg( \frac{\partial f_2}{\partial \tau_c} \bigg)_{\{ T_i \},\{ \alpha_i \} } = 0. 
\end{equation}
$\eta$ only depends on $( {\partial f_1}/{\partial \eta} )_{\{ T_i \}} =0$, and the solution leads to
\begin{equation}
	\eta_{\text{MP}} = 1 - \sqrt{\frac{T_c}{T_h}}.
\end{equation}
It is worth noting that, as shown in the next section, even if the conductivities $\alpha_c$ and $\alpha_h$ depend on the driving protocols, our results Eqs.~\eqref{approx_1} and \eqref{approx_2} can still give a good estimate for $\eta_{\text{MP}}^{(2)}$ and $\eta_{\text{MP}}$.

	\section{Brownian heat engine} \label{sec_5}
	
In this section, we discuss the endoreversible Brownian heat engine which consists of an overdamped Brownian particle trapped in a time-dependent harmonic oscillator potential. We first introduce the Brownian heat engine and give an expression of the cumulants of work output and heat input. Then, we show that the Brownian heat engine satisfies the endoreversible approximation and is compatible with the linear heat transfer laws.

	\subsection{Statistics of work and heat} \label{sec_5a}

The Hamiltonian of the working substance in the Brownian heat engine is given by 
	\begin{equation}
		H(x,t) = V(x,t)=\frac{1}{2}\lambda(t)x^2 ,
	\end{equation}
where $x$ is the position of the Brownian particle and $\lambda(t)$ is the stiffness of the potential which can be controlled externally. There are two external control parameters in our system, $\lambda(t)$ and the water temperature $T(t)$, and they are changed cyclically with the cycle period $\tau$: $\lambda(t + \tau) = \lambda(t)$ and $T(t+\tau)=T(t)$.
	
The time evolution of the phase-space PDF $p(x,t)$ is given by the following Fokker-Planck equation \cite{Seifert_2012}:
	\begin{equation}
		\partial_{t} p(x,t)= \mu \left\{ \partial_x \left[ \left( \partial_x V \right)  p  \right] + k_B T \partial_x^2 p \right\}  \label{F-P} ,
	\end{equation}
where $\mu$ is the mobility of the Brownian particle. When the engine is stable, the PDF is also periodic in time with the period $\tau$: $p(x,t) = p(x,t+\tau)$. Such a periodic solution of Eq.~\eqref{F-P} is \cite{Schmiedl_2007}
	\begin{equation}
		p(x,t)=\frac{1}{\sqrt{2\pi \sigma_x(t)}}\exp{\bigg \{-\frac{x^2}{2\sigma_x(t)}\bigg \}} \label{p}
	\end{equation}
with $\langle x(t) \rangle = 0$ and $\sigma_x(t) \equiv \langle (x - \langle x \rangle)^2 \rangle = \langle x^2(t) \rangle$ whose equation of motion is given by
	\begin{equation}
		\dot{\sigma}_x=-2\mu\lambda \sigma_x + 2\mu k_B T . \label{EOM}
	\end{equation}
Here, the dot represents the derivative with respect to time.
In this model, the correlation function $\phi(t_1,t_2) \equiv \langle x(t_1) x(t_2) \rangle$ is analytically solvable \cite{handbook}, and the solution is given by
\begin{align}
	\phi(t_1,t_2) =& e^{-[f(t_1)+f(t_2)]} 2\mu \\ \notag
	&\times \bigg( \int_0^{\min(t_1,t_2)} e^{2f(t)}T(t) \mathrm{d}t + \frac{\int_0^{\tau}e^{2f(t)}T(t) \mathrm{d}t}{e^{2f(\tau)}-1} \bigg)
\end{align}
with $f(t) \equiv \mu\int_0^t \mathrm{d}t' \lambda(t')$. 
	
The work output through one cycle is given by
	\begin{align}
		W[x(t)|_0^{\tau}] & = - \int_0^{\tau} \frac{\partial V(x,t)}{\partial \lambda}\dot{\lambda}(t) \dd t \notag \\
		& = - \int_0^{\tau} x^2(t) \frac{\dot{\lambda}(t)}{2} \dd t .
	\end{align}
The mean value of work in an infinitesimal time interval $\dd t$ is given by 
\begin{equation}
	\langle \dbar W \rangle = - \sigma_x(t) \dot{\lambda}(t) \dd t /2 . \label{meanW}
\end{equation}
During the heat transfer stroke $ 0 \to 1$ from $t_0$ to $t_1$, the amount of heat flowing into the engine from the hot heat bath is given by
	\begin{align}
		Q_{h}[x(t)|_{t_0}^{t_1}] & = V(x,t_1) - V(x,t_0) + W[x(t)|_{t_0}^{t_1}]  \notag \\
		& = \int_{t_0}^{t_1} \dd t~ x^2(t) \bigg( \delta(t-t_1) - \delta(t-t_0) - \frac{1}{2} \partial_{t} \bigg)\lambda(t) .
	\end{align}
From Eq.~\eqref{meanW} and the first law of thermodynamics, $\E{\dot{H}} = \E{\dot{Q}} - \E{\dot{W}}$, the mean value of the heat transfer rate is given by 
	\begin{equation}
		\langle \dot{Q} \rangle = \dot{\sigma}_x(t) \lambda(t) /2 .\label{heat_rate}
	\end{equation}
	
The work output $W$ and the heat input $Q_h$ are the same functional form as
	\begin{equation}
		\zeta_{\nu}[x(t)|_0^{\tau}] = \int_0^{\tau} \dd t~\frac{K_{\nu}(t) }{2} x^2(t)
	\end{equation} 
with $\nu=1$ for $W$ and $\nu=2$ for $Q_h$. We have $K_1(t) = -\dot{\lambda}(t)$ and $K_2(t) = [\delta(t-t_1)-\delta(t-t_0)-u(t) \partial_{t}]\lambda(t)$, where $u(t) = 1$ when $t_0<t<t_1$ and $u(t)=0$ otherwise.
Since $x(t)$ is a Gaussian process, we can calculate the $n$th order cumulant $\langle \zeta_{\nu}^n \rangle_c$ ($n>1$) using Wick's theorem, and obtain
\begin{align}
	\langle \zeta_{\nu}^n \rangle_c =& \frac{(n-1)!}{2} \int \dd t_1 K_{\nu}(t_1) \int \dd t_2 K_{\nu}(t_2) \ldots \notag \\ 
	&\times \int \dd t_n K_{\nu}(t_n) \prod_{i=1}^{n-1} \phi(t_i,t_{i+1}) \phi(t_n,t_1) . \label{cumulant}
\end{align}
For example, the variance (i.e., the second order cumulant with $n=2$) of work output $W_{t_0}^{(N)}$ in $N$ cycles from $t_0$ to $t_0+N\tau$ is given by
	\begin{equation}
		\langle (\Delta W_{t_0}^{(N)})^2 \rangle =
		\frac{1}{2}\int_{t_0}^{N\tau+t_0} \dd t_1~ \dot{\lambda}(t_1) \int_{t_0}^{N\tau+t_0} \dd t_2~ \dot{\lambda}(t_2)\phi^2(t_1,t_2).
	\end{equation}
Generally speaking, the cumulants of work and heat in finite number of cycles depend on the starting point. The starting-point dependence becomes negligible in the quasistatic case, or when the number of cycles $N$ of the operation is infinite. Furthermore, for finite $\tau$, since the correlations of work and heat between two different cycles are non-negligible, work and heat fluctuations in infinite cycles can be very different from those in a single cycle \cite{xu2021correlationenhanced}. In the present paper, we consider the fluctuations for the latter case.

	\subsection{Endoreversiblility of the Brownian heat engines} \label{sec_5b}

Now, we discuss the endoreversibility of the Brownian heat engines.
We still consider the four-stroke cycles consisting of two heat transfer strokes and two adiabatic strokes. During each adiabatic stroke, the Shannon entropy $S = - k_B \langle \ln p \rangle$ does not change. One way to realize such adiabatic strokes without heat exchange ($Q=0$) is quenching $T$ and $\lambda$ simultaneously \cite{Schmiedl_2007}. Such adiabatic strokes can be implemented in the current experiments. For example, in the experiment of Ref.~\cite{Brownian_Carnot}, the heat engine consists of a charged Brownian particle whose $\lambda$ is controlled by tuning the intensity of the trapping laser and $T$ is controlled by applying a noisy electric force to the charged Brownian particle.
Therefore, $\lambda$ and $T$ can be changed very fast compared to the timescale of the heat exchange between the bath and the working substance.

Since the PDF of the overdamped Brownian particle in our system is always in the canonical distribution given by Eq.~\eqref{p}, we can define the effective temperature $\tilde{T}$ by the width of the PDF as \cite{Schmiedl_2007}
	\begin{equation}
		k_B \tilde{T} (t) = \lambda(t) \sigma_x(t) . \label{Teff}
	\end{equation} 
Thus, the heat capacity defined by the effective temperature is constant:
\begin{equation}
	C_V = \bigg\langle \bigg( \dfrac{\partial H}{\partial \tilde{T} } \bigg)_{\lambda} \bigg\rangle =\dfrac{k_B}{2} .
\end{equation}
From Eqs.~\eqref{EOM}, \eqref{heat_rate}, and \eqref{Teff}, we get the heat transfer rate as
	\begin{equation}
		\langle \dot{Q} \rangle = \frac{1}{2} \lambda \dot{\sigma}_x (t) = \mu \lambda k_B (T - \tilde{T}) . \label{heat_transf}
	\end{equation}
Since the Shannon entropy $S = - k_B \langle \ln p \rangle$ does not change after a cycle, the entropy production $\Sigma$ per cycle defined in Eq.~\eqref{EP} becomes
	\begin{equation}
		\Sigma = - \bigg\langle \int_0^{\tau} \frac{\dot{Q}}{\tilde{T}} \dd t \bigg\rangle = -\frac{1}{2} \oint \dd \ln \sigma_x = 0. \label{endo_Brownian}
	\end{equation}
Therefore, this model satisfies the endoreversible assumption.

According to Eq.~\eqref{heat_transf}, there are two ways to satisfy the linear heat transfer law. In the first way, the control parameter $\lambda$ is fixed during the heat transfer strokes as in the Otto cycle, which gives the Fourier law:
	\begin{equation}
		\langle \dot{Q} \rangle = \alpha (T - \tilde{T} ), \label{Fourier}
	\end{equation}
where $\alpha = k_B \mu \lambda$ is the thermal conductivity. In the second way, $\tilde{T}$ is fixed during the heat transfer strokes as in the Curzon-Ahlborn heat engine \cite{chen2021microscopic}, which gives the Newton law:
	\begin{equation}
		\langle Q \rangle = \alpha \tau (T - \tilde{T} ), \label{Newton}
	\end{equation} 
where $\alpha = (k_B\mu/\tau) \int_0^{\tau} \lambda(t) \mathrm{d}t$ is the time-averaged thermal conductivity. Using the above two ways of the control protocol in our model, we can construct the small Otto and Curzon-Ahlborn engines under the endoreversible condition. In the following, we shall discuss the two examples in detail. 
	
\section{Examples} \label{sec_6}

In this section, we discuss the Brownian Otto engine and Brownian CA engine as examples and examine our main results, the approximate relations \eqref{approx_1} and \eqref{approx_2}. For the both cases, we first derive the maximum power and the efficiency $\eta_{\text{MP}}$ under a specific constraint condition. Then, we calculate $\eta_{\text{MP}}^{(2)}$ and discuss the relation between $\eta_{\text{MP}}$, $\eta_{\text{MP}}^{(2)}$, and $\eta_{CA}$. In addition, for the Brownian Otto engine, we obtain analytical expressions of the cumulant and PDF of work.

	\subsection{Brownian Otto engine} \label{sec_6a}

\begin{figure}[t]
	\includegraphics[width=1 \columnwidth]{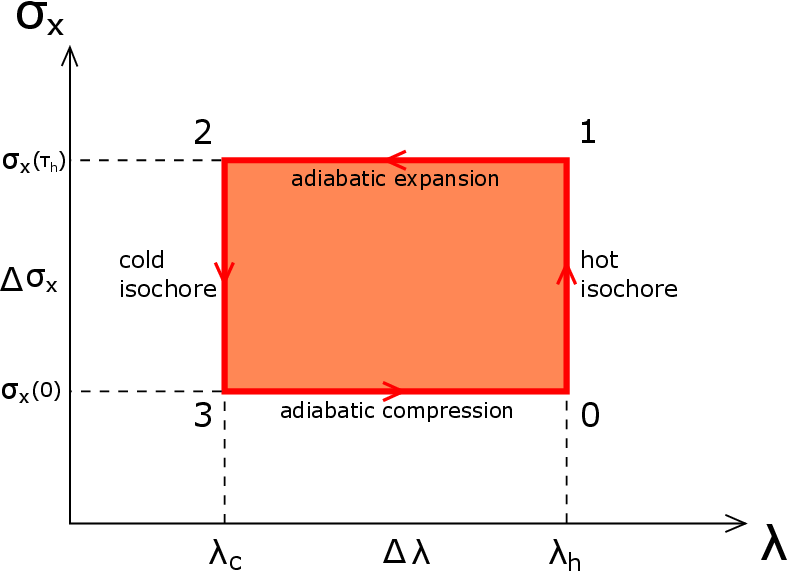} 
	\caption{Otto cycle in the $\sigma_x$-$\lambda$ plane.}
	\label{fig:Otto}
\end{figure}

First, we introduce the Brownian Otto engine which consists of two isochoric and two adiabatic strokes. During the hot (cold) isochoric stroke, the temperature $T$ of the bath and the parameter $\lambda$ are fixed at $T_h$ and $\lambda_h$ ($T_c$ and $\lambda_c$), respectively, for the duration $\tau_h$ ($\tau_c$) with $\lambda_c<\lambda_h$. In the adiabatic strokes, $T$ and $\lambda$ are quenched simultaneously in a way such that the Shannon entropy of the working substance is unchanged \cite{Schmiedl_2007}. We define the time $t_j$ of node $j$ as $t_0 = 0$, $t_1 = t_2 = \tau_h$, and $t_3=t_4 = \tau = \tau_c + \tau_h$. As shown in Fig.~\ref{fig:Otto}, since the variance $\sigma_x$ of the position of the particle does not change during the adiabatic jumps and $\lambda$ is constant during the isochoric strokes, the cycle is a rectangle in the $\sigma_x$-$\lambda$ plane. From Eq.~\eqref{Teff}, we have the effective temperature $\tilde{T}_j$ at node $j$ as $\tilde{T}_0 = \lambda_h \sigma_x(0)$, $\tilde{T}_1=\lambda_h \sigma_x(\tau_h)$, $\tilde{T}_2=\lambda_c \sigma_x(\tau_h)$, and $\tilde{T}_3 = \lambda_c \sigma_x(\tau)= \lambda_c \sigma_x(0)$. For this microscopic endoreversible Otto engine, according to Eq.~\eqref{eta_Teff_theta1}, we have the efficiency:
	\begin{equation}
		\eta = 1 - \frac{\tilde{T}_3}{\tilde{T}_0} = 1-\frac{\lambda_c}{\lambda_h} , \label{eta_Otto}
	\end{equation}
which is determined solely by the ratio $\lambda_c/\lambda_h$. 

	\subsubsection{Maximum power and efficiency} \label{sec_6a1}
Now, we derive the maximum power $P_{\text{max}}$ and the efficiency $\eta_{\text{MP}}$.
It is noted that, for the Otto engine, if the thermal conductivities $\alpha_h$ and $\alpha_c$ are constant, we have $\eta_{\text{MP}} =\eta_{CA}$ as discussed in Sec.~\ref{sec_4}. 
However, in the present microscopic model, the conductivity $\alpha = k_B \mu \lambda$ depends on $\lambda$. From Eq.~\eqref{Power} with $C=k_B/2$ and $\gamma=1$, the power is given by
	\begin{equation}
		P = \frac{k_B\eta[(1-\eta) T_h-T_c]}{(1-\eta)(\tau_h+\tau_c)}\frac{\sinh(\mu (1-\eta)\lambda_h \tau_c)\sinh(\mu \lambda_h \tau_h)}{\sinh(\mu (1-\eta)\lambda_h \tau_c + \mu \lambda_h \tau_h)}. \label{P_Otto}
	\end{equation}
Here, the power is a function of the parameter $\lambda_h$, the efficiency, and the durations for given $T_h$, $T_c$, and $\mu$, i.e., $P = P(\lambda_h,\eta,\tau_h,\tau_c;T_h,T_c,\mu)$. Since $P$ depends on $\lambda_h$ and $\lambda_c$ with the ratio $\lambda_c/\lambda_h=1-\eta$, we should optimize both $\lambda_h$ and $\eta$ for the fully optimized case. However, as shown later, the power increases with $\lambda_h$ so that, if we optimize $\lambda_h$, the maximum power diverges at infinite $\lambda_h$. Since $\lambda$ is finite and can be easily fixed in experiments, the case with optimized $\eta$ for given $\lambda_h$ is more useful compared to the fully optimized case.

\begin{figure}[t]
	\includegraphics[width=1 \columnwidth]{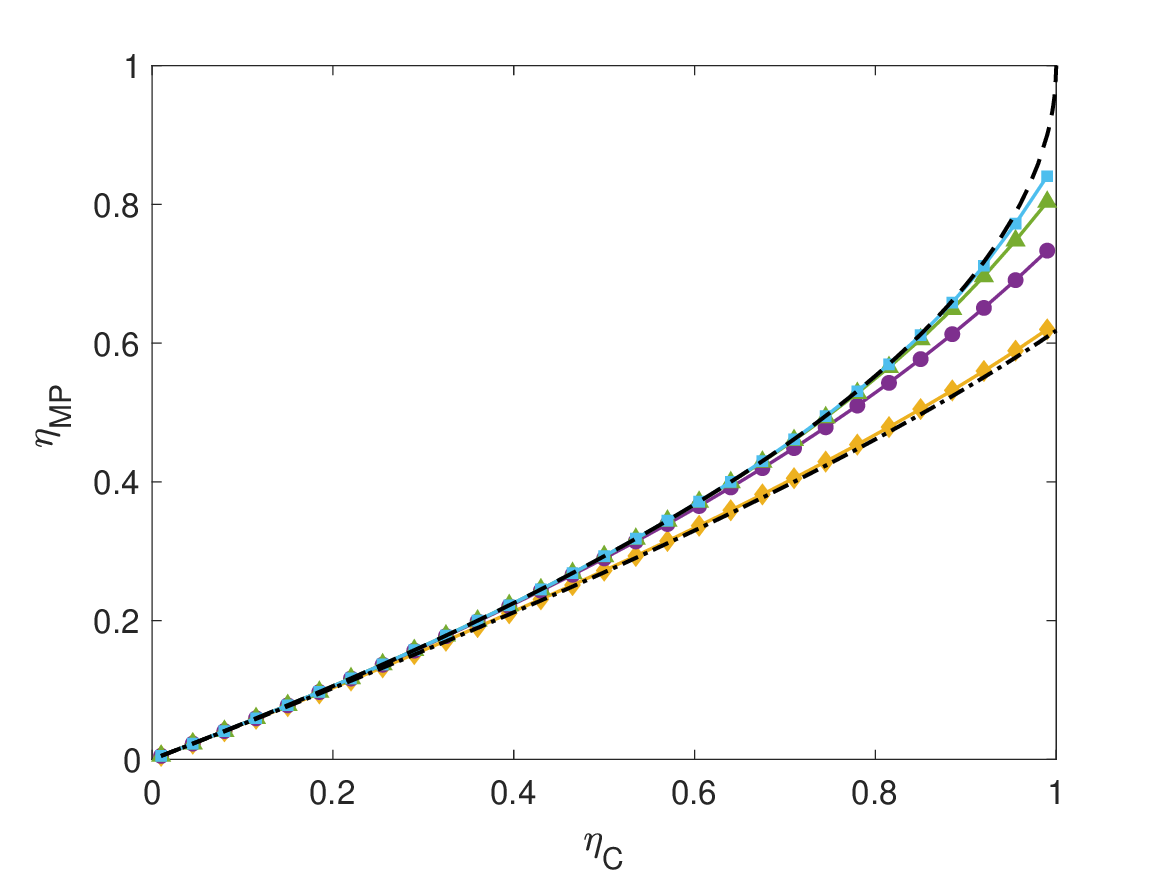} 
	\caption{$\eta_{\text{MP}}$ as a function of $\eta_C$ for various values of $\tau$. In the limit of the large cycle period $\tau/(\mu \lambda_h)^{-1} \to \infty$, we have $\eta_{\text{MP}} = \eta_{CA}$ shown by the black dashed line. In the limit of the small cycle period $\tau/(\mu \lambda_h)^{-1} \to 0$, the resulting $\eta_{\text{MP}}$ is shown by the black dashed-dotted line. The solid lines with symbols show $\eta_{\text{MP}}$ for $\tau/(\mu \lambda_h)^{-1} = 1$ (yellow diamonds), $5$ (purple circles), $9$ (green triangles), and $13$ (cyan squares). }
	\label{fig:EMP_Otto}
\end{figure}

To get the efficiency at maximum power under given $\tau$, $\lambda_h$, $T_h$, $T_c$, and $\mu$, we optimize $\eta$, $\tau_h$, and $\tau_c$ with the constraint,
\begin{equation}
	\tau=\tau_h+\tau_c, \label{tau}
\end{equation}
using the Lagrange multiplier method. With a Lagrange multiplier $\xi$, we define a function $P'$ as
	\begin{equation}
		P'(\eta,\tau_h,\tau_c) \equiv P(\eta,\tau_h,\tau_c) - \xi \cdot (\tau_h+\tau_c).
	\end{equation}
For the maximum power, we have
\begin{equation}
	\bigg( \frac{\partial P'}{\partial \eta} \bigg)_{\tau_h,\tau_c} = \bigg( \frac{\partial P'}{\partial \tau_h} \bigg)_{\eta,\tau_c} = \bigg( \frac{\partial P'}{\partial \tau_c} \bigg)_{\eta,\tau_h} = 0. \label{optim}
\end{equation}
By eliminating $\xi$ from the three equations of Eq.~\eqref{optim}, we get 
\begin{equation}
		(1-\eta)\sinh^2(\tau_h\mu\lambda_h) = \sinh^2[\tau_c(1-\eta)\mu\lambda_h], \label{L1}
\end{equation}
and
\begin{align}
 	\frac{T_c}{T_h} &- (1-\eta)^2 \notag \\
	&+ \frac{\eta[(1-\eta) -\frac{T_c}{T_h} ](1-\eta)\mu\lambda_h\tau_c \sinh(\mu\lambda_h\tau_h)}{\sinh[(1-\eta)\mu\lambda_h\tau_c]\sinh[(1-\eta)\mu\lambda_h\tau_c+\mu\lambda_h\tau_h]} = 0. \label{L2}
\end{align}
The solution of the combined nonlinear equations \eqref{L1} and \eqref{L2} together with the constraint condition \eqref{tau} gives $\eta_{\text{MP}}$ and optimized durations $\tau_h^*$ and $\tau_c^*$ for given $\tau$. Note that, in these equations, $\mu \lambda_h$ always appears in the form of $\mu \lambda_h \tau_i$ for $i = h$ and $c$. We thus define new variables $\tilde{\tau}_i = \mu \lambda_h \tau_i$. Then the solutions $\eta_{\text{MP}}$ and $\tilde{\tau}_i^*$ depend only on the cycle period $\tau$ and the temperature ratio $T_c/T_h = 1-\eta_C$. Figure~\ref{fig:EMP_Otto} shows the deviation of $\eta_{\text{MP}}$ from $\eta_{CA}$ for given values of $\tau$. 

\begin{figure}[t]
	\includegraphics[width=1 \columnwidth]{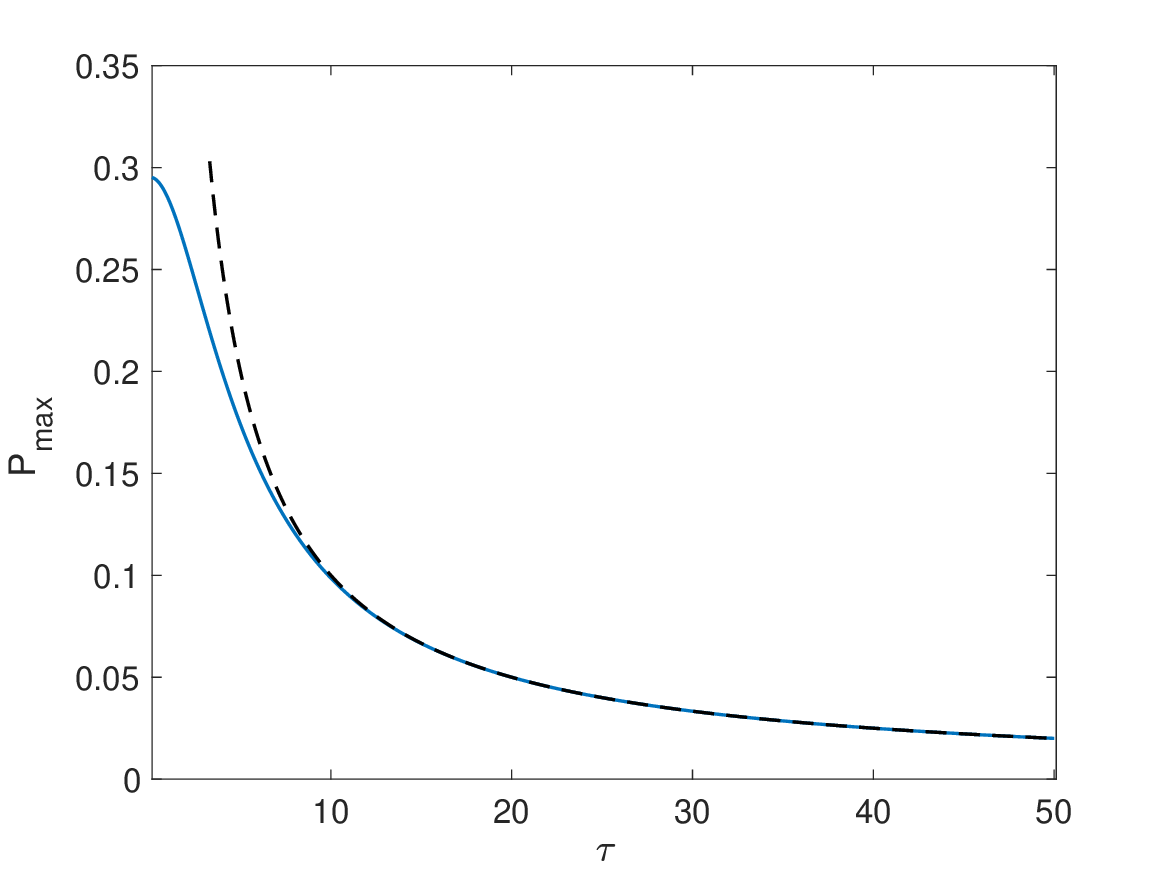} 
	\caption{$P_{\text{max}}$ as a function of $\tau$. We set $T_c/T_h=0.1$. $P_{\text{max}}$ is in units of $\mu \lambda_h k_B T_h \eta_{CA}^2/2$ and $\tau$ is in units of $(\mu\lambda_h)^{-1}$. The dashed line shows $1/\tau$.}
	\label{fig:Pmax_Otto}
\end{figure}

Substituting $\eta_{\text{MP}}$, $\tilde{\tau}_h^*$, and $\tilde{\tau}_c^*$ into Eq.~\eqref{P_Otto}, we get the maximum power $P_{\text{max}}(\tau)$ as a function of the cycle period $\tau$ which is in the form of 
\begin{equation}
	P_{\text{max}}(\tau)=\mu\lambda_h k_B T_h~f(\tau;T_c/T_h) 
\end{equation}
with given $\mu \lambda_h$, $T_h$, and $T_c$. Here, $f(\tau;T_c/T_h)$ is a dimensionless function. Figure~\ref{fig:Pmax_Otto} shows the numerical result of $P_{\text{max}}(\tau)$. Note that $P_{\text{max}}(\tau)$ is maximized when the cycle period $\tau$ approaches zero. To get a physical understanding, let us consider the mean value of work in the $\mathcal{P}$-$\lambda$ plane, where $\mathcal{P}$ is the generalized pressure defined as $\mathcal{P} = \E{\partial H/\partial \lambda} = \sigma_x /2$. Then, as can be seen from Fig.~\ref{fig:Otto}, the mean value of the work output through one cycle is given by $\E{W} = \Delta \lambda \Delta \sigma_x /2$. Thus, the power is proportional to the changing rate of $\sigma_x$: $P \propto \Delta \sigma_x / \tau$. According to Eqs.~\eqref{EOM} and \eqref{Teff}, the changing rate $\dot{\sigma}_x$ is proportional to the temperature difference $T-\tilde{T}$ between the working substance and the bath. Since the temperature difference is finite and takes a maximum value at the beginning of each heat transfer stroke, $\dot{\sigma}_x$ is finite and maximized when $\tau \to 0$. Therefore, the power is also finite and maximized at $\tau \to 0$.

As can be seen from Fig.~\ref{fig:EMP_Otto}, the value of $\eta_{\text{MP}}$ increases as the cycle period and is between the black dashed line and the black dashed-dotted line. Here, the black dashed (dashed-dotted) line shows $\eta_{\text{MP}}$ in the limit of large (small) cycle period. 
For large durations $\tau_h \gg (\mu\lambda_h)^{-1}$ and $\tau_c \gg (\mu\lambda_c)^{-1}$, the third term in the LHS of Eq.~\eqref{L2} is negligible. Thus, we get 
\begin{equation}
	\eta_{\text{MP}}=1-\sqrt{\frac{T_c}{T_h} }=\eta_{CA} 
\end{equation}
from Eq.~\eqref{L2}, and
\begin{equation}
	P_{\text{max}}=k_B T_h \eta_{CA}^2/(2\tau)
\end{equation}
from Eq.~\eqref{P_Otto}. 

For small cycle period $\tau \ll (\mu\lambda_h)^{-1}$, we have $\sinh\tau_i \simeq \tau_i$, and Eqs.~\eqref{L1} and \eqref{L2} become
\begin{equation}
	\eta_{\text{MP}} = 1 - \frac{\tau_h^2}{\tau_c^2} \label{SL1}
\end{equation}
and 
\begin{equation}
	\left( 2\eta_{\text{MP}} - \eta_C \right) \tau_h = [\eta_{\text{MP}}^2 - 2\eta_{\text{MP}} + \eta_C]\tau_c  \label{SL2} ,
\end{equation}
respectively.
Substituting Eq.~\eqref{SL1} into Eq.~\eqref{SL2}, we get
	\begin{equation}
		\eta_{\text{MP}}(1+\sqrt{1-\eta_{\text{MP}}}) = \eta_C . \label{EMP_Otto_tau0}
	\end{equation}
From Eqs.~\eqref{SL1} and \eqref{EMP_Otto_tau0}, the maximum power given by Eq.~\eqref{P_Otto} becomes
\begin{equation}
	P_{\text{max}}= \mu\lambda_h k_B T_h (\eta_C-\eta_{\text{MP}}) \eta_{\text{MP}}^3/\eta_C^2 .
\end{equation}
For $T_c/T_h\simeq 1$ so that $\eta_C \ll 1 $, from Eq.~\eqref{EMP_Otto_tau0} we can expend $\eta_{\text{MP}}$ as
	\begin{equation}
		\eta_{\text{MP}} = \frac{1}{2}\eta_C + \frac{1}{16} \eta_C^2 + \frac{3}{128} \eta_C^3 + O(\eta_C^4) .
	\end{equation}
On the other hand, expanding the CA efficiency $\eta_{CA} = 1-\sqrt{T_c/T_h}$ around $T_c/T_h=1$, we obtain
\begin{equation}
	\eta_{CA} = \frac{1}{2}\eta_C + \frac{1}{8} \eta_C^2 + \frac{3}{16} \eta_C^3 + O(\eta_C^4) .
\end{equation}

In conclusion, 
we have 
\begin{equation}
		\eta_{\text{MP}} \lesssim \eta_{CA} \label{eta_MP_CA_Otto}
\end{equation} 
for the endoreversible small Otto engine. Here, $\eta_{\text{MP}}$ approaches $\eta_{CA}$ as the cycle period increases or $\eta_C$ decreases. 

	\subsubsection{Ratio $\eta_{\textsc{MP}}^{(2)}$ at maximum power} \label{sec_6a2}
Since there is no work output during the heat transfer strokes in the Otto cycle, work and heat are determined solely by the energy change in each stroke:
\begin{equation}
	W = \dfrac{1}{2} (\lambda_h-\lambda_c)[ x(\tau_h)^2 - x(0)^2 ]
\end{equation}
and 
\begin{equation}
	Q_h = \dfrac{1}{2} \lambda_h[ x(\tau_h)^2 - x(0)^2 ] .
\end{equation}
Therefore, the stochastic efficiency $\tilde{\eta} \equiv W/Q_h = 1-\lambda_c/\lambda_h$ becomes deterministic and agrees with the efficiency $\eta = \E{W}/\E{Q_h} = 1-\lambda_c/\lambda_h$. In addition, $\eta^{(2)} \equiv \E{(\Delta{W})^2}/\E{(\Delta{Q_h})^2}$ becomes 
\begin{equation}
	\eta^{(2)} = \bigg( 1 - \frac{\lambda_c}{\lambda_h} \bigg)^2 = \eta^2 .  \label{eta2_eta_Otto}
\end{equation}
It is noted that the relations $\eta = 1-(\lambda_c/\lambda_h) < \eta_C$ and $\eta^{(2)} = \eta^2 < \eta_C^2$ hold for any cycle period within the Otto engine. Thus, from Eqs.~\eqref{eta_MP_CA_Otto} and \eqref{eta2_eta_Otto}, we have 
\begin{equation}
\eta_{\text{MP}}^{(2)} = \eta_{\text{MP}}^2 \lesssim \eta_{CA}^2.  \label{eta2_Otto}
\end{equation}
In addition, since $\eta_{CA} \le \eta_C$, $\eta_C^2$ is still the upper bound of $\eta_{\text{MP}}^{(2)}$ for the endoreversible small Otto engine.

\subsubsection{Cumulant of work} \label{sec_6a3}

The higher order cumulants $\langle W^n \rangle_c$ of work output can be calculated analytically from Eq.~\eqref{cumulant}. Note that the higher order cumulants generally depend on the starting point of the cycle \cite{xu2021correlationenhanced}. For the Otto cycle, the $n$th order cumulant of work output is given by
	\begin{equation}
		\langle W^n \rangle_c = \frac{(n-1)!}{2} \bigg[ (1-c)a^n+(1-c)b^n+c(a+b)^n \bigg]  \label{cumulant_otto}
	\end{equation}
with $a = (\lambda_h-\lambda_c)\sigma_x(\tau_h)$, $b = (\lambda_c-\lambda_h)\sigma_x(0)$, and $c = \phi(\tau_h,\tau)/[\sigma_x(0) \sigma_x(\tau_h)]$ when starting just before the adiabatic expansion; with $a = (\lambda_c-\lambda_h)\sigma_x(0)$, $b =  (\lambda_h-\lambda_c)\sigma_x(\tau_h)$, and $c = \phi(0,\tau_h)/[\sigma_x(0) \sigma_x(\tau_h)]$ when starting just before the adiabatic compression. For some special cases, we can get an analytical expression of the PDF of work $P(W)$. For example, when $a+b=0$, we have 
\begin{equation}
	P(W) = \frac{K_0\left( |W/a| \right)}{\pi |a|} ,
\end{equation}
where $K_0(x)$ is the zeroth order modified Bessel function of the second kind. In addition, for the heat input $Q_h = W/\eta$, all the cumulants are given by $\langle Q_h^n \rangle_c = \eta^{-n}\langle W^n \rangle_c$. The PDFs of work and heat given by the modified Bessel function have been obtained also in other similar systems \cite{PDF_1,PDF_2,PDF_exp,PDF_3,PDF_4,PDF_5,PDF_6,PDF_7}.

	\subsection{Brownian Curzon-Ahlborn heat engine}  \label{sec_6b}

For the small Curzon-Ahlborn heat engine, we assume that the working substance follows the Carnot cycle which consists of two adiabatic jumps and two isothermal strokes. Suppose we have the adiabatic expantion (compression) at $t=\tau_h$ ($t=0$) and the cycle period $\tau=\tau_c+\tau_h$. During the hot (cold) isothermal strokes, both the water temperature $T_h$ ($T_c$) and the effective temperature of the working substance $\tilde{T}_h$ ($\tilde{T}_c$) are constant. As a consequence, $ \dot{\sigma}_x = 2\mu k_B ( T_{i} - \tilde{T}_i )$ with $i = h,c$ given by Eq.~\eqref{EOM} is constant. Thus, the effective temperatures $\tilde{T}_h$ and $\tilde{T}_c$ are given by
	\begin{align}
		\tilde{T}_h &= T_h - \frac{ \sigma_x(\tau_h) - \sigma_x(0) }{2\mu k_B\tau_h} , \label{Tih}  \\
		\tilde{T}_c &= T_c + \frac{ \sigma_x(\tau_h) - \sigma_x(0) }{2\mu k_B\tau_c}.  \label{Tic} 
	\end{align} 
From the Newton law \eqref{Newton}, the thermal conductivity is given by $\alpha_i = \mu \tilde{T}_i [\ln\sigma_x(\tau_h) - \ln\sigma_x(0)]/[\sigma_x(\tau_h) - \sigma_x(0) ]$. If the conductivity is constant, the efficiency at maximum power is the CA efficiency. However, here $\alpha_i(\tilde{T}_i;\sigma_x,\mu)$ is no longer constant, but depends on the effective temperature of the working substance.

\subsubsection{Efficiency at maximum power} \label{sec_6b1}

\begin{figure}[t]
	\includegraphics[width=1 \columnwidth]{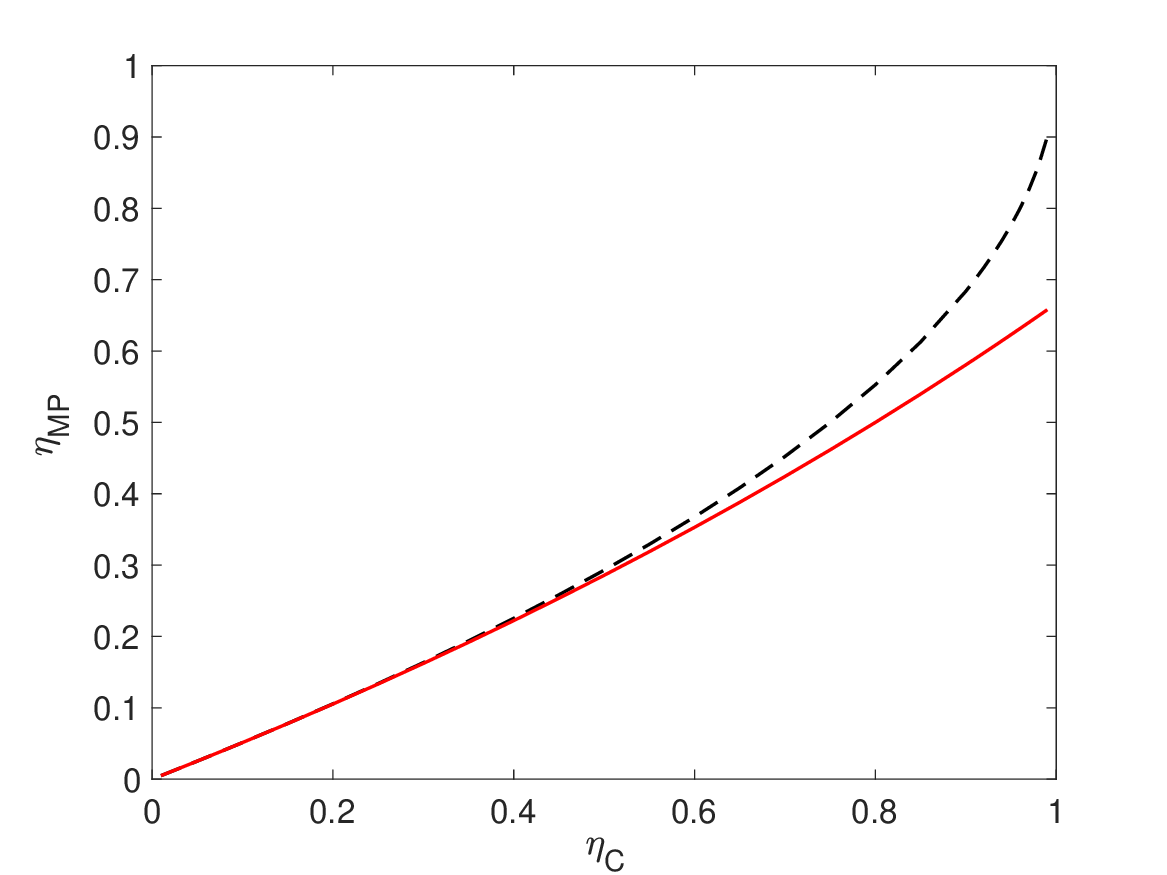} 
	\caption{$\eta_{\text{MP}}$ for the microscopic CA engine as a function of $\eta_C$. The dashed line shows $\eta_{CA}$.}
	\label{fig:EMP_CA}
\end{figure}

Now, we derive $\eta_{\text{MP}}$ of the microscopic CA engine. The power is given by
	\begin{align}
		P & \equiv \frac{\langle W \rangle}{\tau} \notag \\
		&= -\frac{1}{2\tau} \int \sigma_x \mathrm{d}\lambda  \notag\\
		& = -\frac{1}{2\tau} \int \sigma_x \mathrm{d}\bigg( \frac{k_B \tilde{T}_i}{\sigma_x} \bigg)  \notag\\
		& = \frac{k_B}{2\tau} (\tilde{T}_h - \tilde{T}_c) \ln \frac{\sigma_x(\tau_h)}{\sigma_x(0)}  \notag\\
		& = \frac{k_B}{2} \ln \frac{\sigma_x(\tau_h)}{\sigma_x(0)}\bigg[ \frac{T_h - T_c}{\tau_c+\tau_h} - \frac{\sigma_x(\tau_h) - \sigma_x(0) }{2\mu k_B \tau_c\tau_h} \bigg] .
	\end{align}
By maximizing the power with respect to $\tau_c$ and $\tau_h$ under given $T_h$, $T_c$, $\mu$, $\sigma_x(0)$, and $\sigma_x(\tau_h)$, we get $\tau_c=\tau_h=\tau_{\text{MP}} /2$ with 
	\begin{equation}
		\tau_{\text{MP}}=\frac{4[\sigma_x(\tau_h) - \sigma_x(0) ] }{\mu k_B (T_h - T_c)}. \label{tau_MP_CA}
	\end{equation}
Substituting $\tau_{\text{MP}}$ into Eqs.~\eqref{Tih} and \eqref{Tic}, we get the effective temperatures: $\tilde{T}_h = (3/4)T_h + (1/4)T_c$ and $\tilde{T}_c = (1/4)T_h + (3/4)T_c$.
Then $\eta_{\text{MP}}$ is given by
\begin{equation}
	\eta_{\text{MP}} = 1 - \frac{\tilde{T}_c}{\tilde{T}_h} = \frac{2(T_h - T_c)}{3T_h+T_c} = \frac{\eta_C}{2-\eta_C/2}. \label{EMP_CA}
\end{equation}
This result agrees with that of Ref.~\cite{Schmiedl_2007} for the finite-time Brownian heat engines at maximum power. As shown in Fig.~\ref{fig:EMP_CA}, the approximate relation $\eta_{\text{MP}} \simeq \eta_{CA}$ holds when $\eta_C \ll 1 $.

	\subsubsection{Ratio  $\eta_{\textsc{MP}}^{(2)}$ at maximum power} \label{sec_6b2}
\begin{figure}[t]
	\includegraphics[width=1 \columnwidth]{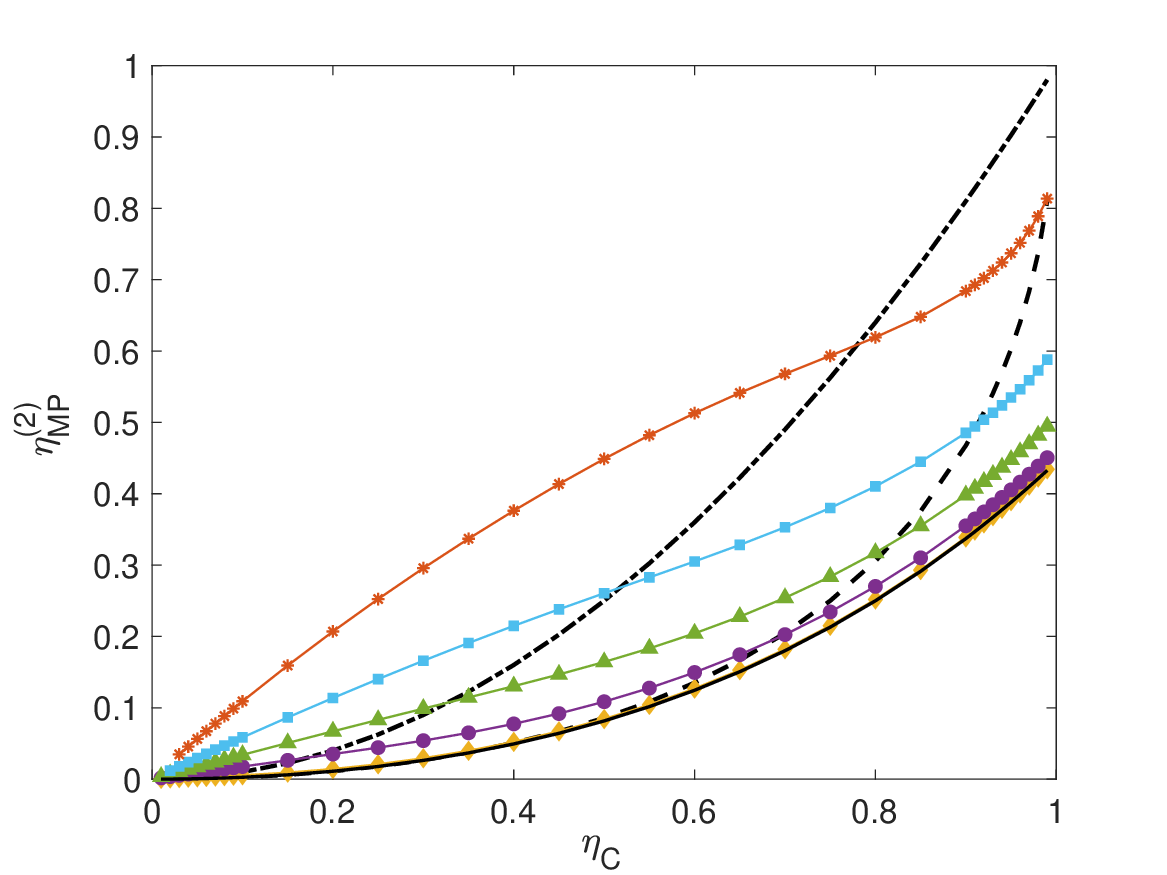} 
	\caption{$\eta_{\text{MP}}^{(2)}$ as a function of $\eta_C$ for various values of $\sigma_x(0)/\sigma_x(\tau_h)$. Solid lines with symbols show $\eta_{\text{MP}}^{(2)}$ for $\sigma_x(0)/\sigma_x(\tau_h) = 0.1$ (orange asterisks), $0.3$ (cyan squares), $0.5$ (green triangles), $0.7$ (purple circles),  and $0.9$ (yellow diamonds). The black dashed-dotted line shows $\eta_C^2$, the black dashed line shows $\eta_{CA}^2$, and the black solid line shows $\eta_{\text{MP}}^2$ with $\eta_{\text{MP}}$ given by Eq.~\eqref{EMP_CA}. } 
	\label{fig:eta2_CA}
\end{figure}

Numerically, we find that $\eta^{(2)}_{\text{MP}}$ for the CA engine depends only on the ratio $\sigma_x(0)/\sigma_x(\tau_h)$ and $\eta_C$. As shown in Fig.~\ref{fig:eta2_CA}, we have $\eta^{(2)}_{\text{MP}} \simeq \eta_{\text{MP}}^2$ for $\sigma_x(0)/\sigma_x(\tau_h) \simeq 1$. That is because when $\sigma_x(0) \simeq \sigma_x(\tau_h)$, $\lambda = k_B \tilde{T} /\sigma_x$ is almost unchanged during the isothermal strokes. Then, the isothermal strokes become isochoric and the CA engine becomes identical to the Otto engine. In this case, work is extracted only in the reversible adiabatic strokes and we have $\eta^{(2)}_{\text{MP}} = \eta_{\text{MP}}^2$ as discussed in the example of the Otto engine. Since $\eta_{\text{MP}}$ is close to $\eta_{CA}$ for the CA engine, we have $\eta^{(2)}_{\text{MP}  } \simeq \eta_{CA}^2$ when $\sigma_x(0)$ is close to $\sigma_x(\tau_h)$. The situation here is very different from that in Sec.~\ref{sec_4}. In Sec.~\ref{sec_4}, we assume that the heat conductivities are constant, and the cycle period is sufficiently large such that the effect of the correlation of work is negligible. Here, from Eq.~\eqref{tau_MP_CA}, we find that the condition $\sigma_x(0)/\sigma_x(\tau_h) \simeq 1$ gives a very small cycle period. Therefore, in the case of the CA engine, the approximate relation $\eta^{(2)}_{\text{MP}  } \simeq \eta_{CA}^2$ holds even when the cycle period is small and the heat conductivities are not constant. 

In addition, the relation $\eta^{(2)}_{\text{MP}  } \simeq \eta_{CA}^2 < \eta_C^2$ also indicates that the upper bound $\eta_C^2$, is still applicable to the CA engine with $\sigma_x(0)/\sigma_x(\tau_h) \simeq 1$. However, as shown in Fig.~\ref{fig:eta2_CA}, $\eta_{\text{MP}}^{(2)}$ could exceed $\eta_C^2$ for the CA engine with small $\eta_C$ and small $\sigma_x(0)/\sigma_x(\tau_h)$.

	\section{Summary and conclusion}
In summary, we have studied the ratio between the variances of work output and heat input, $\eta^{(2)}$, for endoreversible small heat engines. We have found that the ratio $\eta_{\text{MP}}^{(2)}$ at maximum power is equal or close to the square of the Curzon-Ahlborn (CA) efficiency, $\eta_{\text{MP}}^{(2)} \simeq \eta_{CA}^2$, for endoreversible small heat engines. 

Endoreversible small heat engines have the working substance following a reversible cycle and the finite irreversible heat flow causing the finite-time effect. Thus, we separated our discussion into two parts: (i) we first considered $\eta^{(2)}$ for reversible small heat engines in the quasistatic limit in Sec.~\ref{sec_3}, and (ii) we discussed the ratio $\eta_{\text{MP}}^{(2)}$ at maximum power for the endoreversible small heat engines in Sec.~\ref{sec_4}. We considered a class of four-stroke heat engines operating with two heat baths with the temperatures $T_h$ and $T_c$ consisting of two adiabatic strokes and two heat transfer strokes with constant heat capacities $C_h$ and $C_c$. In part (i), the ratio $\theta = C_h/C_c$ is crucial for the relation between $\eta^{(2)}$ and $\eta^2$. (a) For $\theta < 1$, $\eta^2$ gives a lower bound of $\eta^{(2)}$; (b) for $\theta>1$, $\eta^2$ does not give the lower bound of $\eta^{(2)}$ any more, and $\eta^{(2)}$ could be larger or smaller than $\eta^2$; (c) in the typical case of $\theta=1$ (e.g., the Otto cycle, the Brayton cycle, and the Carnot cycle), we obtained $\eta^{(2)}=\eta^2$. In part (ii), we considered the typical case of $\theta=1$. We assumed the heat transfer following the Fourier law or the Newton law. For the constant heat conductivities, when the cycle period is much larger than the correlation time of work and heat, we obtained $\eta_{\text{MP}}^{(2)} = \eta_{\text{MP}}^2 = \eta_{CA}^2= (1-\sqrt{T_c/T_h})^2$ in Sec.~\ref{sec_4}. 
Then, in the last two sections (Secs.~\ref{sec_5} and \ref{sec_6}), we have verified that $\eta_{CA}^2$ gives a good estimate of $\eta_{\text{MP}}^{(2)}$, i.e., $\eta_{\text{MP}}^{(2)} \simeq \eta_{CA}^2$, for the endoreversible Brownian heat engines. In Sec.~\ref{sec_5}, we introduced the endoreversible Brownian heat engine and gave an expression of the cumulant of work and heat. In Sec.~\ref{sec_6}, by taking the two type of the Brownian heat engine cycle as examples, we have shown that the approximate relation $\eta^{(2)}_{\text{MP}  } \simeq \eta_{CA}^2$ holds even when the cycle period is smaller than the correlation time of work and heat and the heat conductivities are not constant. We discussed (a) the Brownian Otto engine in Sec.~\ref{sec_6a} and (b) the Brownian CA engine in Sec.~\ref{sec_6b}.
(a) For the Otto engine, since $\eta^{(2)} = \eta^2$ ($\eta \le \eta_C$) always holds and $\eta_{\text{MP}}$ is close to $\eta_{CA}$, we have $\eta_{\text{MP}}^{(2)} \simeq \eta_{CA}^2$. (b) For the CA engine, we have found that $\eta_{\text{MP}}^{(2)} \simeq \eta_{CA}^2$ holds when the work in the isothermal strokes is small. In this case, the CA engine is close to the Otto engine, and the work is mostly done through the reversible adiabatic strokes. In addition, since $\eta_{CA}^2 \le \eta_C^2$, the upper bound of $\eta^{(2)}$ in the quasistatic limit, $\eta_C^2$, is applicable to some endoreversible small heat engines at maximum power when the relation $\eta_{\text{MP}}^{(2)} \simeq \eta_{CA}^2$ holds. 

Our result, $\eta_{\text{MP}}^{(2)} \simeq \eta_{CA}^2$, resembles the relation $\eta_{\text{MP}} \simeq \eta_{CA}$, i.e., as the CA efficiency gives a good estimate of the efficiency at maximum power for various kinds of finite-time heat engines, $\eta_{CA}^2$ also gives a good estimate of $\eta_{\text{MP}}^{(2)} $ for various finite-time small heat engines. Since the smaller $\eta^{(2)}$ means more stable work output converted from the fluctuating heat input, $\eta_{\text{MP}}^{(2)} \simeq \eta_{CA}^2$ also suggests a trade-off between the efficiency and the stability of finite-time heat engines at maximum power. 
In the future, finding an upper and a lower bound of $\eta_{\text{MP}}^{(2)}$ and minimizing the fluctuation of work output for the finite-time heat engines are interesting open problems. Also, thermodynamic geometry \cite{Crook_geo,Van_geo,Brandner_geo,Izumida_2204.00807,watanabe2021finitetime,Miller_geo,Frim_geo} is a good method which leads to further studies on the properties of $\eta^{(2)}$.

	\section*{Acknowledgement}
	
We thank Kosuke Ito for helpful discussions and comments. G.~W. is supported by NSF of China (Grant No. 11975199), by the Zhejiang Provincial Natural Science Foundation Key Project (Grant No. LZ19A050001).

	\bibliography{eta2_ref}

\begin{thebibliography}{78}%
\makeatletter
\providecommand \@ifxundefined [1]{%
 \@ifx{#1\undefined}
}%
\providecommand \@ifnum [1]{%
 \ifnum #1\expandafter \@firstoftwo
 \else \expandafter \@secondoftwo
 \fi
}%
\providecommand \@ifx [1]{%
 \ifx #1\expandafter \@firstoftwo
 \else \expandafter \@secondoftwo
 \fi
}%
\providecommand \natexlab [1]{#1}%
\providecommand \enquote  [1]{``#1''}%
\providecommand \bibnamefont  [1]{#1}%
\providecommand \bibfnamefont [1]{#1}%
\providecommand \citenamefont [1]{#1}%
\providecommand \href@noop [0]{\@secondoftwo}%
\providecommand \href [0]{\begingroup \@sanitize@url \@href}%
\providecommand \@href[1]{\@@startlink{#1}\@@href}%
\providecommand \@@href[1]{\endgroup#1\@@endlink}%
\providecommand \@sanitize@url [0]{\catcode `\\12\catcode `\$12\catcode
  `\&12\catcode `\#12\catcode `\^12\catcode `\_12\catcode `\%12\relax}%
\providecommand \@@startlink[1]{}%
\providecommand \@@endlink[0]{}%
\providecommand \url  [0]{\begingroup\@sanitize@url \@url }%
\providecommand \@url [1]{\endgroup\@href {#1}{\urlprefix }}%
\providecommand \urlprefix  [0]{URL }%
\providecommand \Eprint [0]{\href }%
\providecommand \doibase [0]{https://doi.org/}%
\providecommand \selectlanguage [0]{\@gobble}%
\providecommand \bibinfo  [0]{\@secondoftwo}%
\providecommand \bibfield  [0]{\@secondoftwo}%
\providecommand \translation [1]{[#1]}%
\providecommand \BibitemOpen [0]{}%
\providecommand \bibitemStop [0]{}%
\providecommand \bibitemNoStop [0]{.\EOS\space}%
\providecommand \EOS [0]{\spacefactor3000\relax}%
\providecommand \BibitemShut  [1]{\csname bibitem#1\endcsname}%
\let\auto@bib@innerbib\@empty
\bibitem [{\citenamefont {Hugel}\ \emph {et~al.}(2002)\citenamefont {Hugel},
  \citenamefont {Holland}, \citenamefont {Cattani}, \citenamefont {Moroder},
  \citenamefont {Seitz},\ and\ \citenamefont {Gaub}}]{Hugel}%
  \BibitemOpen
  \bibfield  {author} {\bibinfo {author} {\bibfnamefont {T.}~\bibnamefont
  {Hugel}}, \bibinfo {author} {\bibfnamefont {N.~B.}\ \bibnamefont {Holland}},
  \bibinfo {author} {\bibfnamefont {A.}~\bibnamefont {Cattani}}, \bibinfo
  {author} {\bibfnamefont {L.}~\bibnamefont {Moroder}}, \bibinfo {author}
  {\bibfnamefont {M.}~\bibnamefont {Seitz}},\ and\ \bibinfo {author}
  {\bibfnamefont {H.~E.}\ \bibnamefont {Gaub}},\ }\bibfield  {title} {\bibinfo
  {title} {Single-{M}olecule {O}ptomechanical {C}ycle},\ }\href
  {https://doi.org/10.1126/science.1069856} {\bibfield  {journal} {\bibinfo
  {journal} {Science}\ }\textbf {\bibinfo {volume} {296}},\ \bibinfo {pages}
  {1103} (\bibinfo {year} {2002})}\BibitemShut {NoStop}%
\bibitem [{\citenamefont {Steeneken}\ \emph {et~al.}(2011)\citenamefont
  {Steeneken}, \citenamefont {Le~Phan}, \citenamefont {Goossens}, \citenamefont
  {Koops}, \citenamefont {Brom}, \citenamefont {van~der Avoort},\ and\
  \citenamefont {van Beek}}]{Steeneken}%
  \BibitemOpen
  \bibfield  {author} {\bibinfo {author} {\bibfnamefont {P.~G.}\ \bibnamefont
  {Steeneken}}, \bibinfo {author} {\bibfnamefont {K.}~\bibnamefont {Le~Phan}},
  \bibinfo {author} {\bibfnamefont {M.~J.}\ \bibnamefont {Goossens}}, \bibinfo
  {author} {\bibfnamefont {G.~E.~J.}\ \bibnamefont {Koops}}, \bibinfo {author}
  {\bibfnamefont {G.~J. A.~M.}\ \bibnamefont {Brom}}, \bibinfo {author}
  {\bibfnamefont {C.}~\bibnamefont {van~der Avoort}},\ and\ \bibinfo {author}
  {\bibfnamefont {J.~T.~M.}\ \bibnamefont {van Beek}},\ }\bibfield  {title}
  {\bibinfo {title} {Piezoresistive heat engine and refrigerator},\ }\href
  {https://doi.org/10.1038/nphys1871} {\bibfield  {journal} {\bibinfo
  {journal} {Nat. Phys.}\ }\textbf {\bibinfo {volume} {7}},\ \bibinfo {pages}
  {354} (\bibinfo {year} {2011})}\BibitemShut {NoStop}%
\bibitem [{\citenamefont {Toyabe}\ \emph {et~al.}(2010)\citenamefont {Toyabe},
  \citenamefont {Sagawa}, \citenamefont {Ueda}, \citenamefont {Muneyuki},\ and\
  \citenamefont {Sano}}]{Shoichi}%
  \BibitemOpen
  \bibfield  {author} {\bibinfo {author} {\bibfnamefont {S.}~\bibnamefont
  {Toyabe}}, \bibinfo {author} {\bibfnamefont {T.}~\bibnamefont {Sagawa}},
  \bibinfo {author} {\bibfnamefont {M.}~\bibnamefont {Ueda}}, \bibinfo {author}
  {\bibfnamefont {E.}~\bibnamefont {Muneyuki}},\ and\ \bibinfo {author}
  {\bibfnamefont {M.}~\bibnamefont {Sano}},\ }\bibfield  {title} {\bibinfo
  {title} {Experimental demonstration of information-to-energy conversion and
  validation of the generalized {J}arzynski equality},\ }\href
  {https://doi.org/10.1038/nphys1821} {\bibfield  {journal} {\bibinfo
  {journal} {Nat. Phys.}\ }\textbf {\bibinfo {volume} {6}},\ \bibinfo {pages}
  {988} (\bibinfo {year} {2010})}\BibitemShut {NoStop}%
\bibitem [{\citenamefont {Blickle}\ and\ \citenamefont
  {Bechinger}(2012)}]{Blickle_experiment}%
  \BibitemOpen
  \bibfield  {author} {\bibinfo {author} {\bibfnamefont {V.}~\bibnamefont
  {Blickle}}\ and\ \bibinfo {author} {\bibfnamefont {C.}~\bibnamefont
  {Bechinger}},\ }\bibfield  {title} {\bibinfo {title} {Realization of a
  micrometre-sized stochastic heat engine},\ }\href
  {https://www.nature.com/articles/nphys2163} {\bibfield  {journal} {\bibinfo
  {journal} {Nat. Phys.}\ }\textbf {\bibinfo {volume} {8}},\ \bibinfo {pages}
  {143} (\bibinfo {year} {2012})}\BibitemShut {NoStop}%
\bibitem [{\citenamefont {Mart{\'{i}}nez}\ \emph {et~al.}(2016)\citenamefont
  {Mart{\'{i}}nez}, \citenamefont {Rold{\'{a}}n}, \citenamefont {Dinis},
  \citenamefont {Petrov}, \citenamefont {Parrondo},\ and\ \citenamefont
  {Rica}}]{Brownian_Carnot}%
  \BibitemOpen
  \bibfield  {author} {\bibinfo {author} {\bibfnamefont {I.~A.}\ \bibnamefont
  {Mart{\'{i}}nez}}, \bibinfo {author} {\bibfnamefont {{\'{E}}.}~\bibnamefont
  {Rold{\'{a}}n}}, \bibinfo {author} {\bibfnamefont {L.}~\bibnamefont {Dinis}},
  \bibinfo {author} {\bibfnamefont {D.}~\bibnamefont {Petrov}}, \bibinfo
  {author} {\bibfnamefont {J.~M.~R.}\ \bibnamefont {Parrondo}},\ and\ \bibinfo
  {author} {\bibfnamefont {R.~A.}\ \bibnamefont {Rica}},\ }\bibfield  {title}
  {\bibinfo {title} {Brownian {C}arnot engine},\ }\href
  {https://www.nature.com/articles/nphys3518} {\bibfield  {journal} {\bibinfo
  {journal} {Nat. Phys.}\ }\textbf {\bibinfo {volume} {12}},\ \bibinfo {pages}
  {67} (\bibinfo {year} {2016})}\BibitemShut {NoStop}%
\bibitem [{\citenamefont {Krishnamurthy}\ \emph {et~al.}(2016)\citenamefont
  {Krishnamurthy}, \citenamefont {Ghosh}, \citenamefont {Chatterji},
  \citenamefont {Ganapathy},\ and\ \citenamefont {Sood}}]{Krishnamurthy}%
  \BibitemOpen
  \bibfield  {author} {\bibinfo {author} {\bibfnamefont {S.}~\bibnamefont
  {Krishnamurthy}}, \bibinfo {author} {\bibfnamefont {S.}~\bibnamefont
  {Ghosh}}, \bibinfo {author} {\bibfnamefont {D.}~\bibnamefont {Chatterji}},
  \bibinfo {author} {\bibfnamefont {R.}~\bibnamefont {Ganapathy}},\ and\
  \bibinfo {author} {\bibfnamefont {A.~K.}\ \bibnamefont {Sood}},\ }\bibfield
  {title} {\bibinfo {title} {A micrometre-sized heat engine operating between
  bacterial reservoirs},\ }\href {https://www.nature.com/articles/nphys3870}
  {\bibfield  {journal} {\bibinfo  {journal} {Nat. Phys.}\ }\textbf {\bibinfo
  {volume} {12}},\ \bibinfo {pages} {1134} (\bibinfo {year}
  {2016})}\BibitemShut {NoStop}%
\bibitem [{\citenamefont {Argun}\ \emph {et~al.}(2017)\citenamefont {Argun},
  \citenamefont {Soni}, \citenamefont {Dabelow}, \citenamefont {Bo},
  \citenamefont {Pesce}, \citenamefont {Eichhorn},\ and\ \citenamefont
  {Volpe}}]{Argun}%
  \BibitemOpen
  \bibfield  {author} {\bibinfo {author} {\bibfnamefont {A.}~\bibnamefont
  {Argun}}, \bibinfo {author} {\bibfnamefont {J.}~\bibnamefont {Soni}},
  \bibinfo {author} {\bibfnamefont {L.}~\bibnamefont {Dabelow}}, \bibinfo
  {author} {\bibfnamefont {S.}~\bibnamefont {Bo}}, \bibinfo {author}
  {\bibfnamefont {G.}~\bibnamefont {Pesce}}, \bibinfo {author} {\bibfnamefont
  {R.}~\bibnamefont {Eichhorn}},\ and\ \bibinfo {author} {\bibfnamefont
  {G.}~\bibnamefont {Volpe}},\ }\bibfield  {title} {\bibinfo {title}
  {Experimental realization of a minimal microscopic heat engine},\ }\href
  {https://doi.org/10.1103/PhysRevE.96.052106} {\bibfield  {journal} {\bibinfo
  {journal} {Phys. Rev. E}\ }\textbf {\bibinfo {volume} {96}},\ \bibinfo
  {pages} {052106} (\bibinfo {year} {2017})}\BibitemShut {NoStop}%
\bibitem [{\citenamefont {Mart\'{i}nez}\ \emph {et~al.}(2017)\citenamefont
  {Mart\'{i}nez}, \citenamefont {Rold\'{a}n}, \citenamefont {Dinis},\ and\
  \citenamefont {Rica}}]{Colloidal_heat_engines}%
  \BibitemOpen
  \bibfield  {author} {\bibinfo {author} {\bibfnamefont {I.~A.}\ \bibnamefont
  {Mart\'{i}nez}}, \bibinfo {author} {\bibfnamefont {{\'{E}}.}~\bibnamefont
  {Rold\'{a}n}}, \bibinfo {author} {\bibfnamefont {L.}~\bibnamefont {Dinis}},\
  and\ \bibinfo {author} {\bibfnamefont {R.~A.}\ \bibnamefont {Rica}},\
  }\bibfield  {title} {\bibinfo {title} {Colloidal heat engines: a review},\
  }\href {https://doi.org/10.1039/C6SM00923A} {\bibfield  {journal} {\bibinfo
  {journal} {Soft Matter}\ }\textbf {\bibinfo {volume} {13}},\ \bibinfo {pages}
  {22} (\bibinfo {year} {2017})}\BibitemShut {NoStop}%
\bibitem [{\citenamefont {Ciliberto}(2017)}]{Exp_review}%
  \BibitemOpen
  \bibfield  {author} {\bibinfo {author} {\bibfnamefont {S.}~\bibnamefont
  {Ciliberto}},\ }\bibfield  {title} {\bibinfo {title} {Experiments in
  {S}tochastic {T}hermodynamics: {S}hort {H}istory and {P}erspectives},\ }\href
  {https://doi.org/10.1103/PhysRevX.7.021051} {\bibfield  {journal} {\bibinfo
  {journal} {Phys. Rev. X}\ }\textbf {\bibinfo {volume} {7}},\ \bibinfo {pages}
  {021051} (\bibinfo {year} {2017})}\BibitemShut {NoStop}%
\bibitem [{\citenamefont {Erbas-Cakmak}\ \emph {et~al.}(2015)\citenamefont
  {Erbas-Cakmak}, \citenamefont {Leigh}, \citenamefont {McTernan},\ and\
  \citenamefont {Nussbaumer}}]{Exp_molecular}%
  \BibitemOpen
  \bibfield  {author} {\bibinfo {author} {\bibfnamefont {S.}~\bibnamefont
  {Erbas-Cakmak}}, \bibinfo {author} {\bibfnamefont {D.~A.}\ \bibnamefont
  {Leigh}}, \bibinfo {author} {\bibfnamefont {C.~T.}\ \bibnamefont
  {McTernan}},\ and\ \bibinfo {author} {\bibfnamefont {A.~L.}\ \bibnamefont
  {Nussbaumer}},\ }\bibfield  {title} {\bibinfo {title} {Artificial {M}olecular
  {M}achines},\ }\href {https://doi.org/10.1021/acs.chemrev.5b00146} {\bibfield
   {journal} {\bibinfo  {journal} {Chem. Rev.}\ }\textbf {\bibinfo {volume}
  {115}},\ \bibinfo {pages} {10081} (\bibinfo {year} {2015})}\BibitemShut
  {NoStop}%
\bibitem [{\citenamefont {Sekimoto}(2010)}]{sekimoto_book}%
  \BibitemOpen
  \bibfield  {author} {\bibinfo {author} {\bibfnamefont {K.}~\bibnamefont
  {Sekimoto}},\ }\href@noop {} {\emph {\bibinfo {title} {Stochastic
  Energetics}}}\ (\bibinfo  {publisher} {Springer, Berlin, Heidelberg},\
  \bibinfo {year} {2010})\BibitemShut {NoStop}%
\bibitem [{\citenamefont {Seifert}(2012)}]{Seifert_2012}%
  \BibitemOpen
  \bibfield  {author} {\bibinfo {author} {\bibfnamefont {U.}~\bibnamefont
  {Seifert}},\ }\bibfield  {title} {\bibinfo {title} {Stochastic
  thermodynamics, fluctuation theorems and molecular machines},\ }\href
  {https://doi.org/10.1088/0034-4885/75/12/126001} {\bibfield  {journal}
  {\bibinfo  {journal} {Rep. Prog. Phys.}\ }\textbf {\bibinfo {volume} {75}},\
  \bibinfo {pages} {126001} (\bibinfo {year} {2012})}\BibitemShut {NoStop}%
\bibitem [{\citenamefont {Seifert}(2019)}]{Seifert_2019}%
  \BibitemOpen
  \bibfield  {author} {\bibinfo {author} {\bibfnamefont {U.}~\bibnamefont
  {Seifert}},\ }\bibfield  {title} {\bibinfo {title} {From {S}tochastic
  {T}hermodynamics to {T}hermodynamic {I}nference},\ }\href
  {https://doi.org/10.1146/annurev-conmatphys-031218-013554} {\bibfield
  {journal} {\bibinfo  {journal} {Annu. Rev. Condens. Matter Phys.}\ }\textbf
  {\bibinfo {volume} {10}},\ \bibinfo {pages} {171} (\bibinfo {year}
  {2019})}\BibitemShut {NoStop}%
\bibitem [{\citenamefont {Jarzynski}(2011)}]{Jarzynski_review}%
  \BibitemOpen
  \bibfield  {author} {\bibinfo {author} {\bibfnamefont {C.}~\bibnamefont
  {Jarzynski}},\ }\bibfield  {title} {\bibinfo {title} {Equalities and
  {I}nequalities: {I}rreversibility and the {S}econd {L}aw of {T}hermodynamics
  at the {N}anoscale},\ }\href
  {https://doi.org/10.1146/annurev-conmatphys-062910-140506} {\bibfield
  {journal} {\bibinfo  {journal} {Annu. Rev. Condens. Matter Phys.}\ }\textbf
  {\bibinfo {volume} {2}},\ \bibinfo {pages} {329} (\bibinfo {year}
  {2011})}\BibitemShut {NoStop}%
\bibitem [{\citenamefont {Bustamante}\ \emph {et~al.}(2005)\citenamefont
  {Bustamante}, \citenamefont {Liphardt},\ and\ \citenamefont
  {Ritort}}]{Noneq_smallsys}%
  \BibitemOpen
  \bibfield  {author} {\bibinfo {author} {\bibfnamefont {C.}~\bibnamefont
  {Bustamante}}, \bibinfo {author} {\bibfnamefont {J.}~\bibnamefont
  {Liphardt}},\ and\ \bibinfo {author} {\bibfnamefont {F.}~\bibnamefont
  {Ritort}},\ }\bibfield  {title} {\bibinfo {title} {The {N}onequilibrium
  {T}hermodynamics of {S}mall {S}ystems},\ }\href
  {https://doi.org/10.1063/1.2012462} {\bibfield  {journal} {\bibinfo
  {journal} {Phys. Today}\ }\textbf {\bibinfo {volume} {58}},\ \bibinfo {pages}
  {43} (\bibinfo {year} {2005})}\BibitemShut {NoStop}%
\bibitem [{\citenamefont {Ciliberto}\ \emph {et~al.}(2010)\citenamefont
  {Ciliberto}, \citenamefont {Joubaud},\ and\ \citenamefont
  {Petrosyan}}]{Ciliberto_fluct}%
  \BibitemOpen
  \bibfield  {author} {\bibinfo {author} {\bibfnamefont {S.}~\bibnamefont
  {Ciliberto}}, \bibinfo {author} {\bibfnamefont {S.}~\bibnamefont {Joubaud}},\
  and\ \bibinfo {author} {\bibfnamefont {A.}~\bibnamefont {Petrosyan}},\
  }\bibfield  {title} {\bibinfo {title} {Fluctuations in out-of-equilibrium
  systems: from theory to experiment},\ }\href
  {https://doi.org/10.1088/1742-5468/2010/12/p12003} {\bibfield  {journal}
  {\bibinfo  {journal} {J. Stat. Mech.}\ }\textbf {\bibinfo {volume} {2010}},\
  \bibinfo {pages} {P12003} (\bibinfo {year} {2010})}\BibitemShut {NoStop}%
\bibitem [{\citenamefont {Sinitsyn}(2011)}]{Sinitsyn_2011}%
  \BibitemOpen
  \bibfield  {author} {\bibinfo {author} {\bibfnamefont {N.~A.}\ \bibnamefont
  {Sinitsyn}},\ }\bibfield  {title} {\bibinfo {title} {Fluctuation relation for
  heat engines},\ }\href {https://doi.org/10.1088/1751-8113/44/40/405001}
  {\bibfield  {journal} {\bibinfo  {journal} {J. Phys. A: Math. Theor.}\
  }\textbf {\bibinfo {volume} {44}},\ \bibinfo {pages} {405001} (\bibinfo
  {year} {2011})}\BibitemShut {NoStop}%
\bibitem [{\citenamefont {Lahiri}\ \emph {et~al.}(2012)\citenamefont {Lahiri},
  \citenamefont {Rana},\ and\ \citenamefont {Jayannavar}}]{Lahiri_2012}%
  \BibitemOpen
  \bibfield  {author} {\bibinfo {author} {\bibfnamefont {S.}~\bibnamefont
  {Lahiri}}, \bibinfo {author} {\bibfnamefont {S.}~\bibnamefont {Rana}},\ and\
  \bibinfo {author} {\bibfnamefont {A.~M.}\ \bibnamefont {Jayannavar}},\
  }\bibfield  {title} {\bibinfo {title} {Fluctuation relations for heat engines
  in time-periodic steady states},\ }\href
  {https://doi.org/10.1088/1751-8113/45/46/465001} {\bibfield  {journal}
  {\bibinfo  {journal} {J. Phys. A: Math. Theor.}\ }\textbf {\bibinfo {volume}
  {45}},\ \bibinfo {pages} {465001} (\bibinfo {year} {2012})}\BibitemShut
  {NoStop}%
\bibitem [{\citenamefont {Campisi}(2014)}]{Campisi_2014}%
  \BibitemOpen
  \bibfield  {author} {\bibinfo {author} {\bibfnamefont {M.}~\bibnamefont
  {Campisi}},\ }\bibfield  {title} {\bibinfo {title} {Fluctuation relation for
  quantum heat engines and refrigerators},\ }\href
  {https://doi.org/10.1088/1751-8113/47/24/245001} {\bibfield  {journal}
  {\bibinfo  {journal} {J. Phys. A: Math. Theor.}\ }\textbf {\bibinfo {volume}
  {47}},\ \bibinfo {pages} {245001} (\bibinfo {year} {2014})}\BibitemShut
  {NoStop}%
\bibitem [{\citenamefont {Rana}\ \emph {et~al.}(2014)\citenamefont {Rana},
  \citenamefont {Pal}, \citenamefont {Saha},\ and\ \citenamefont
  {Jayannavar}}]{Single-particle_HE}%
  \BibitemOpen
  \bibfield  {author} {\bibinfo {author} {\bibfnamefont {S.}~\bibnamefont
  {Rana}}, \bibinfo {author} {\bibfnamefont {P.~S.}\ \bibnamefont {Pal}},
  \bibinfo {author} {\bibfnamefont {A.}~\bibnamefont {Saha}},\ and\ \bibinfo
  {author} {\bibfnamefont {A.~M.}\ \bibnamefont {Jayannavar}},\ }\bibfield
  {title} {\bibinfo {title} {Single-particle stochastic heat engine},\ }\href
  {https://doi.org/10.1103/PhysRevE.90.042146} {\bibfield  {journal} {\bibinfo
  {journal} {Phys. Rev. E}\ }\textbf {\bibinfo {volume} {90}},\ \bibinfo
  {pages} {042146} (\bibinfo {year} {2014})}\BibitemShut {NoStop}%
\bibitem [{\citenamefont {Zheng}\ and\ \citenamefont
  {Poletti}(2014)}]{Otto_powerlaw}%
  \BibitemOpen
  \bibfield  {author} {\bibinfo {author} {\bibfnamefont {Y.}~\bibnamefont
  {Zheng}}\ and\ \bibinfo {author} {\bibfnamefont {D.}~\bibnamefont
  {Poletti}},\ }\bibfield  {title} {\bibinfo {title} {Work and efficiency of
  quantum {O}tto cycles in power-law trapping potentials},\ }\href
  {https://doi.org/10.1103/PhysRevE.90.012145} {\bibfield  {journal} {\bibinfo
  {journal} {Phys. Rev. E}\ }\textbf {\bibinfo {volume} {90}},\ \bibinfo
  {pages} {012145} (\bibinfo {year} {2014})}\BibitemShut {NoStop}%
\bibitem [{\citenamefont {Ito}\ \emph {et~al.}(2019)\citenamefont {Ito},
  \citenamefont {Jiang},\ and\ \citenamefont {Watanabe}}]{ito2019universal}%
  \BibitemOpen
  \bibfield  {author} {\bibinfo {author} {\bibfnamefont {K.}~\bibnamefont
  {Ito}}, \bibinfo {author} {\bibfnamefont {C.}~\bibnamefont {Jiang}},\ and\
  \bibinfo {author} {\bibfnamefont {G.}~\bibnamefont {Watanabe}},\ }\href@noop
  {} {\bibinfo {title} {Universal {B}ounds for {F}luctuations in {S}mall {H}eat
  {E}ngines}} (\bibinfo {year} {2019}),\ \Eprint
  {https://arxiv.org/abs/1910.08096} {arXiv:1910.08096 [cond-mat.stat-mech]}
  \BibitemShut {NoStop}%
\bibitem [{\citenamefont {Saryal}\ and\ \citenamefont
  {Agarwalla}(2021)}]{Saryal_fluct_qOtto}%
  \BibitemOpen
  \bibfield  {author} {\bibinfo {author} {\bibfnamefont {S.}~\bibnamefont
  {Saryal}}\ and\ \bibinfo {author} {\bibfnamefont {B.~K.}\ \bibnamefont
  {Agarwalla}},\ }\bibfield  {title} {\bibinfo {title} {Bounds on fluctuations
  for finite-time quantum {O}tto cycle},\ }\href
  {https://doi.org/10.1103/PhysRevE.103.L060103} {\bibfield  {journal}
  {\bibinfo  {journal} {Phys. Rev. E}\ }\textbf {\bibinfo {volume} {103}},\
  \bibinfo {pages} {L060103} (\bibinfo {year} {2021})}\BibitemShut {NoStop}%
\bibitem [{\citenamefont {Saryal}\ \emph {et~al.}(2022)\citenamefont {Saryal},
  \citenamefont {Mohanta},\ and\ \citenamefont
  {Agarwalla}}]{saryal2021universal}%
  \BibitemOpen
  \bibfield  {author} {\bibinfo {author} {\bibfnamefont {S.}~\bibnamefont
  {Saryal}}, \bibinfo {author} {\bibfnamefont {S.}~\bibnamefont {Mohanta}},\
  and\ \bibinfo {author} {\bibfnamefont {B.~K.}\ \bibnamefont {Agarwalla}},\
  }\bibfield  {title} {\bibinfo {title} {Bounds on fluctuations for machines
  with broken time-reversal symmetry: {A} linear response study},\ }\href
  {https://doi.org/10.1103/PhysRevE.105.024129} {\bibfield  {journal} {\bibinfo
   {journal} {Phys. Rev. E}\ }\textbf {\bibinfo {volume} {105}},\ \bibinfo
  {pages} {024129} (\bibinfo {year} {2022})}\BibitemShut {NoStop}%
\bibitem [{\citenamefont {Miller}\ and\ \citenamefont
  {Mehboudi}(2020)}]{Miller_geo}%
  \BibitemOpen
  \bibfield  {author} {\bibinfo {author} {\bibfnamefont {H.~J.~D.}\
  \bibnamefont {Miller}}\ and\ \bibinfo {author} {\bibfnamefont
  {M.}~\bibnamefont {Mehboudi}},\ }\bibfield  {title} {\bibinfo {title}
  {Geometry of {W}ork {F}luctuations versus {E}fficiency in {M}icroscopic
  {T}hermal {M}achines},\ }\href
  {https://doi.org/10.1103/PhysRevLett.125.260602} {\bibfield  {journal}
  {\bibinfo  {journal} {Phys. Rev. Lett.}\ }\textbf {\bibinfo {volume} {125}},\
  \bibinfo {pages} {260602} (\bibinfo {year} {2020})}\BibitemShut {NoStop}%
\bibitem [{\citenamefont {Watanabe}\ and\ \citenamefont
  {Minami}(2022)}]{watanabe2021finitetime}%
  \BibitemOpen
  \bibfield  {author} {\bibinfo {author} {\bibfnamefont {G.}~\bibnamefont
  {Watanabe}}\ and\ \bibinfo {author} {\bibfnamefont {Y.}~\bibnamefont
  {Minami}},\ }\bibfield  {title} {\bibinfo {title} {Finite-time thermodynamics
  of fluctuations in microscopic heat engines},\ }\href
  {https://doi.org/10.1103/PhysRevResearch.4.L012008} {\bibfield  {journal}
  {\bibinfo  {journal} {Phys. Rev. Research}\ }\textbf {\bibinfo {volume}
  {4}},\ \bibinfo {pages} {L012008} (\bibinfo {year} {2022})}\BibitemShut
  {NoStop}%
\bibitem [{\citenamefont {Holubec}\ and\ \citenamefont
  {Ryabov}(2021)}]{Holubec_2021}%
  \BibitemOpen
  \bibfield  {author} {\bibinfo {author} {\bibfnamefont {V.}~\bibnamefont
  {Holubec}}\ and\ \bibinfo {author} {\bibfnamefont {A.}~\bibnamefont
  {Ryabov}},\ }\bibfield  {title} {\bibinfo {title} {Fluctuations in heat
  engines},\ }\href {https://doi.org/10.1088/1751-8121/ac3aac} {\bibfield
  {journal} {\bibinfo  {journal} {J. Phys. A: Math. Theor.}\ }\textbf {\bibinfo
  {volume} {55}},\ \bibinfo {pages} {013001} (\bibinfo {year}
  {2021})}\BibitemShut {NoStop}%
\bibitem [{\citenamefont {Chen}\ \emph {et~al.}(2022)\citenamefont {Chen},
  \citenamefont {Chen}, \citenamefont {Fei},\ and\ \citenamefont
  {Quan}}]{chen2021microscopic}%
  \BibitemOpen
  \bibfield  {author} {\bibinfo {author} {\bibfnamefont {Y.~H.}\ \bibnamefont
  {Chen}}, \bibinfo {author} {\bibfnamefont {J.-F.}\ \bibnamefont {Chen}},
  \bibinfo {author} {\bibfnamefont {Z.}~\bibnamefont {Fei}},\ and\ \bibinfo
  {author} {\bibfnamefont {H.~T.}\ \bibnamefont {Quan}},\ }\bibfield  {title}
  {\bibinfo {title} {{Microscopic theory of the Curzon-Ahlborn heat engine
  based on a Brownian particle}},\ }\href
  {https://doi.org/10.1103/PhysRevE.106.024105} {\bibfield  {journal} {\bibinfo
   {journal} {Phys. Rev. E}\ }\textbf {\bibinfo {volume} {106}},\ \bibinfo
  {pages} {024105} (\bibinfo {year} {2022})}\BibitemShut {NoStop}%
\bibitem [{\citenamefont {Kwon}\ \emph {et~al.}(2013)\citenamefont {Kwon},
  \citenamefont {Noh},\ and\ \citenamefont {Park}}]{work_fluct}%
  \BibitemOpen
  \bibfield  {author} {\bibinfo {author} {\bibfnamefont {C.}~\bibnamefont
  {Kwon}}, \bibinfo {author} {\bibfnamefont {J.~D.}\ \bibnamefont {Noh}},\ and\
  \bibinfo {author} {\bibfnamefont {H.}~\bibnamefont {Park}},\ }\bibfield
  {title} {\bibinfo {title} {Work fluctuations in a time-dependent harmonic
  potential: {R}igorous results beyond the overdamped limit},\ }\href
  {https://doi.org/10.1103/PhysRevE.88.062102} {\bibfield  {journal} {\bibinfo
  {journal} {Phys. Rev. E}\ }\textbf {\bibinfo {volume} {88}},\ \bibinfo
  {pages} {062102} (\bibinfo {year} {2013})}\BibitemShut {NoStop}%
\bibitem [{\citenamefont {Salazar}(2020)}]{Work_distribution}%
  \BibitemOpen
  \bibfield  {author} {\bibinfo {author} {\bibfnamefont {D.~S.~P.}\
  \bibnamefont {Salazar}},\ }\bibfield  {title} {\bibinfo {title} {Work
  distribution in thermal processes},\ }\href
  {https://doi.org/10.1103/PhysRevE.101.030101} {\bibfield  {journal} {\bibinfo
   {journal} {Phys. Rev. E}\ }\textbf {\bibinfo {volume} {101}},\ \bibinfo
  {pages} {030101(R)} (\bibinfo {year} {2020})}\BibitemShut {NoStop}%
\bibitem [{\citenamefont {Verley}\ \emph
  {et~al.}(2014{\natexlab{a}})\citenamefont {Verley}, \citenamefont {Esposito},
  \citenamefont {Willaert},\ and\ \citenamefont {Van~den
  Broeck}}]{stocastic_efficiency}%
  \BibitemOpen
  \bibfield  {author} {\bibinfo {author} {\bibfnamefont {G.}~\bibnamefont
  {Verley}}, \bibinfo {author} {\bibfnamefont {M.}~\bibnamefont {Esposito}},
  \bibinfo {author} {\bibfnamefont {T.}~\bibnamefont {Willaert}},\ and\
  \bibinfo {author} {\bibfnamefont {C.}~\bibnamefont {Van~den Broeck}},\
  }\bibfield  {title} {\bibinfo {title} {The unlikely {C}arnot efficiency},\
  }\href {https://doi.org/10.1038/ncomms5721} {\bibfield  {journal} {\bibinfo
  {journal} {Nat. Commun.}\ }\textbf {\bibinfo {volume} {5}},\ \bibinfo {pages}
  {4721} (\bibinfo {year} {2014}{\natexlab{a}})}\BibitemShut {NoStop}%
\bibitem [{\citenamefont {Verley}\ \emph
  {et~al.}(2014{\natexlab{b}})\citenamefont {Verley}, \citenamefont {Willaert},
  \citenamefont {Van~den Broeck},\ and\ \citenamefont
  {Esposito}}]{stocastic_efficiency_2}%
  \BibitemOpen
  \bibfield  {author} {\bibinfo {author} {\bibfnamefont {G.}~\bibnamefont
  {Verley}}, \bibinfo {author} {\bibfnamefont {T.}~\bibnamefont {Willaert}},
  \bibinfo {author} {\bibfnamefont {C.}~\bibnamefont {Van~den Broeck}},\ and\
  \bibinfo {author} {\bibfnamefont {M.}~\bibnamefont {Esposito}},\ }\bibfield
  {title} {\bibinfo {title} {Universal theory of efficiency fluctuations},\
  }\href {https://link.aps.org/doi/10.1103/PhysRevE.90.052145} {\bibfield
  {journal} {\bibinfo  {journal} {Phys. Rev. E}\ }\textbf {\bibinfo {volume}
  {90}},\ \bibinfo {pages} {052145} (\bibinfo {year}
  {2014}{\natexlab{b}})}\BibitemShut {NoStop}%
\bibitem [{\citenamefont {Polettini}\ \emph {et~al.}(2015)\citenamefont
  {Polettini}, \citenamefont {Verley},\ and\ \citenamefont
  {Esposito}}]{Stoch_eta_3}%
  \BibitemOpen
  \bibfield  {author} {\bibinfo {author} {\bibfnamefont {M.}~\bibnamefont
  {Polettini}}, \bibinfo {author} {\bibfnamefont {G.}~\bibnamefont {Verley}},\
  and\ \bibinfo {author} {\bibfnamefont {M.}~\bibnamefont {Esposito}},\
  }\bibfield  {title} {\bibinfo {title} {Efficiency {S}tatistics at {A}ll
  {T}imes: {C}arnot {L}imit at {F}inite {P}ower},\ }\href
  {https://doi.org/10.1103/PhysRevLett.114.050601} {\bibfield  {journal}
  {\bibinfo  {journal} {Phys. Rev. Lett.}\ }\textbf {\bibinfo {volume} {114}},\
  \bibinfo {pages} {050601} (\bibinfo {year} {2015})}\BibitemShut {NoStop}%
\bibitem [{\citenamefont {Jiang}\ \emph {et~al.}(2015)\citenamefont {Jiang},
  \citenamefont {Agarwalla},\ and\ \citenamefont {Segal}}]{SE_4}%
  \BibitemOpen
  \bibfield  {author} {\bibinfo {author} {\bibfnamefont {J.-H.}\ \bibnamefont
  {Jiang}}, \bibinfo {author} {\bibfnamefont {B.~K.}\ \bibnamefont
  {Agarwalla}},\ and\ \bibinfo {author} {\bibfnamefont {D.}~\bibnamefont
  {Segal}},\ }\bibfield  {title} {\bibinfo {title} {Efficiency {S}tatistics and
  {B}ounds for {S}ystems with {B}roken {T}ime-{R}eversal {S}ymmetry},\ }\href
  {https://doi.org/10.1103/PhysRevLett.115.040601} {\bibfield  {journal}
  {\bibinfo  {journal} {Phys. Rev. Lett.}\ }\textbf {\bibinfo {volume} {115}},\
  \bibinfo {pages} {040601} (\bibinfo {year} {2015})}\BibitemShut {NoStop}%
\bibitem [{\citenamefont {Proesmans}\ \emph {et~al.}(2015)\citenamefont
  {Proesmans}, \citenamefont {Cleuren},\ and\ \citenamefont {Van~den
  Broeck}}]{SE_5}%
  \BibitemOpen
  \bibfield  {author} {\bibinfo {author} {\bibfnamefont {K.}~\bibnamefont
  {Proesmans}}, \bibinfo {author} {\bibfnamefont {B.}~\bibnamefont {Cleuren}},\
  and\ \bibinfo {author} {\bibfnamefont {C.}~\bibnamefont {Van~den Broeck}},\
  }\bibfield  {title} {\bibinfo {title} {Stochastic efficiency for effusion as
  a thermal engine},\ }\href {https://doi.org/10.1209/0295-5075/109/20004}
  {\bibfield  {journal} {\bibinfo  {journal} {EPL}\ }\textbf {\bibinfo {volume}
  {109}},\ \bibinfo {pages} {20004} (\bibinfo {year} {2015})}\BibitemShut
  {NoStop}%
\bibitem [{\citenamefont {Fischer}\ \emph {et~al.}(2018)\citenamefont
  {Fischer}, \citenamefont {Pietzonka},\ and\ \citenamefont {Seifert}}]{SE_6}%
  \BibitemOpen
  \bibfield  {author} {\bibinfo {author} {\bibfnamefont {L.~P.}\ \bibnamefont
  {Fischer}}, \bibinfo {author} {\bibfnamefont {P.}~\bibnamefont {Pietzonka}},\
  and\ \bibinfo {author} {\bibfnamefont {U.}~\bibnamefont {Seifert}},\
  }\bibfield  {title} {\bibinfo {title} {Large deviation function for a driven
  underdamped particle in a periodic potential},\ }\href
  {https://doi.org/10.1103/PhysRevE.97.022143} {\bibfield  {journal} {\bibinfo
  {journal} {Phys. Rev. E}\ }\textbf {\bibinfo {volume} {97}},\ \bibinfo
  {pages} {022143} (\bibinfo {year} {2018})}\BibitemShut {NoStop}%
\bibitem [{\citenamefont {Manikandan}\ \emph {et~al.}(2019)\citenamefont
  {Manikandan}, \citenamefont {Dabelow}, \citenamefont {Eichhorn},\ and\
  \citenamefont {Krishnamurthy}}]{SE_7}%
  \BibitemOpen
  \bibfield  {author} {\bibinfo {author} {\bibfnamefont {S.~K.}\ \bibnamefont
  {Manikandan}}, \bibinfo {author} {\bibfnamefont {L.}~\bibnamefont {Dabelow}},
  \bibinfo {author} {\bibfnamefont {R.}~\bibnamefont {Eichhorn}},\ and\
  \bibinfo {author} {\bibfnamefont {S.}~\bibnamefont {Krishnamurthy}},\
  }\bibfield  {title} {\bibinfo {title} {Efficiency {F}luctuations in
  {M}icroscopic {M}achines},\ }\href
  {https://doi.org/10.1103/PhysRevLett.122.140601} {\bibfield  {journal}
  {\bibinfo  {journal} {Phys. Rev. Lett.}\ }\textbf {\bibinfo {volume} {122}},\
  \bibinfo {pages} {140601} (\bibinfo {year} {2019})}\BibitemShut {NoStop}%
\bibitem [{\citenamefont {Barato}\ and\ \citenamefont
  {Seifert}(2015)}]{original_TUR}%
  \BibitemOpen
  \bibfield  {author} {\bibinfo {author} {\bibfnamefont {A.~C.}\ \bibnamefont
  {Barato}}\ and\ \bibinfo {author} {\bibfnamefont {U.}~\bibnamefont
  {Seifert}},\ }\bibfield  {title} {\bibinfo {title} {Thermodynamic
  {U}ncertainty {R}elation for {B}iomolecular {P}rocesses},\ }\href
  {https://doi.org/10.1103/PhysRevLett.114.158101} {\bibfield  {journal}
  {\bibinfo  {journal} {Phys. Rev. Lett.}\ }\textbf {\bibinfo {volume} {114}},\
  \bibinfo {pages} {158101} (\bibinfo {year} {2015})}\BibitemShut {NoStop}%
\bibitem [{\citenamefont {Gingrich}\ \emph {et~al.}(2016)\citenamefont
  {Gingrich}, \citenamefont {Horowitz}, \citenamefont {Perunov},\ and\
  \citenamefont {England}}]{proof_ori_TUR}%
  \BibitemOpen
  \bibfield  {author} {\bibinfo {author} {\bibfnamefont {T.~R.}\ \bibnamefont
  {Gingrich}}, \bibinfo {author} {\bibfnamefont {J.~M.}\ \bibnamefont
  {Horowitz}}, \bibinfo {author} {\bibfnamefont {N.}~\bibnamefont {Perunov}},\
  and\ \bibinfo {author} {\bibfnamefont {J.~L.}\ \bibnamefont {England}},\
  }\bibfield  {title} {\bibinfo {title} {Dissipation {B}ounds {A}ll
  {S}teady-{S}tate {C}urrent {F}luctuations},\ }\href
  {https://doi.org/10.1103/PhysRevLett.116.120601} {\bibfield  {journal}
  {\bibinfo  {journal} {Phys. Rev. Lett.}\ }\textbf {\bibinfo {volume} {116}},\
  \bibinfo {pages} {120601} (\bibinfo {year} {2016})}\BibitemShut {NoStop}%
\bibitem [{\citenamefont {Horowitz}\ and\ \citenamefont
  {Gingrich}(2020)}]{review_TUR}%
  \BibitemOpen
  \bibfield  {author} {\bibinfo {author} {\bibfnamefont {J.}~\bibnamefont
  {Horowitz}}\ and\ \bibinfo {author} {\bibfnamefont {T.}~\bibnamefont
  {Gingrich}},\ }\bibfield  {title} {\bibinfo {title} {Thermodynamic
  uncertainty relations constrain non-equilibrium fluctuations},\ }\href
  {https://doi.org/10.1038/s41567-019-0702-6} {\bibfield  {journal} {\bibinfo
  {journal} {Nat. Phys.}\ }\textbf {\bibinfo {volume} {16}},\ \bibinfo {pages}
  {15} (\bibinfo {year} {2020})}\BibitemShut {NoStop}%
\bibitem [{\citenamefont {Pietzonka}\ and\ \citenamefont
  {Seifert}(2018)}]{Pietzonka_Trade-Off}%
  \BibitemOpen
  \bibfield  {author} {\bibinfo {author} {\bibfnamefont {P.}~\bibnamefont
  {Pietzonka}}\ and\ \bibinfo {author} {\bibfnamefont {U.}~\bibnamefont
  {Seifert}},\ }\bibfield  {title} {\bibinfo {title} {Universal {T}rade-{O}ff
  between {P}ower, {E}fficiency, and {C}onstancy in {S}teady-{S}tate {H}eat
  {E}ngines},\ }\href {https://doi.org/10.1103/PhysRevLett.120.190602}
  {\bibfield  {journal} {\bibinfo  {journal} {Phys. Rev. Lett.}\ }\textbf
  {\bibinfo {volume} {120}},\ \bibinfo {pages} {190602} (\bibinfo {year}
  {2018})}\BibitemShut {NoStop}%
\bibitem [{\citenamefont {Holubec}\ and\ \citenamefont
  {Ryabov}(2018)}]{Holubec_CTUR}%
  \BibitemOpen
  \bibfield  {author} {\bibinfo {author} {\bibfnamefont {V.}~\bibnamefont
  {Holubec}}\ and\ \bibinfo {author} {\bibfnamefont {A.}~\bibnamefont
  {Ryabov}},\ }\bibfield  {title} {\bibinfo {title} {Cycling {T}ames {P}ower
  {F}luctuations near {O}ptimum {E}fficiency},\ }\href
  {https://doi.org/10.1103/PhysRevLett.121.120601} {\bibfield  {journal}
  {\bibinfo  {journal} {Phys. Rev. Lett.}\ }\textbf {\bibinfo {volume} {121}},\
  \bibinfo {pages} {120601} (\bibinfo {year} {2018})}\BibitemShut {NoStop}%
\bibitem [{\citenamefont {Barato}\ \emph {et~al.}(2018)\citenamefont {Barato},
  \citenamefont {Chetrite}, \citenamefont {Faggionato},\ and\ \citenamefont
  {Gabrielli}}]{modified_EP_1}%
  \BibitemOpen
  \bibfield  {author} {\bibinfo {author} {\bibfnamefont {A.~C.}\ \bibnamefont
  {Barato}}, \bibinfo {author} {\bibfnamefont {R.}~\bibnamefont {Chetrite}},
  \bibinfo {author} {\bibfnamefont {A.}~\bibnamefont {Faggionato}},\ and\
  \bibinfo {author} {\bibfnamefont {D.}~\bibnamefont {Gabrielli}},\ }\bibfield
  {title} {\bibinfo {title} {Bounds on current fluctuations in periodically
  driven systems},\ }\href
  {https://iopscience.iop.org/article/10.1088/1367-2630/aae512} {\bibfield
  {journal} {\bibinfo  {journal} {New J. Phys.}\ }\textbf {\bibinfo {volume}
  {20}},\ \bibinfo {pages} {103023} (\bibinfo {year} {2018})}\BibitemShut
  {NoStop}%
\bibitem [{\citenamefont {Koyuk}\ \emph {et~al.}(2018)\citenamefont {Koyuk},
  \citenamefont {Seifert},\ and\ \citenamefont {Pietzonka}}]{modified_EP_2}%
  \BibitemOpen
  \bibfield  {author} {\bibinfo {author} {\bibfnamefont {T.}~\bibnamefont
  {Koyuk}}, \bibinfo {author} {\bibfnamefont {U.}~\bibnamefont {Seifert}},\
  and\ \bibinfo {author} {\bibfnamefont {P.}~\bibnamefont {Pietzonka}},\
  }\bibfield  {title} {\bibinfo {title} {A generalization of the thermodynamic
  uncertainty relation to periodically driven systems},\ }\href
  {https://iopscience.iop.org/article/10.1088/1751-8121/aaeec4} {\bibfield
  {journal} {\bibinfo  {journal} {J. Phys. A: Math. Theor.}\ }\textbf {\bibinfo
  {volume} {52}},\ \bibinfo {pages} {02LT02} (\bibinfo {year}
  {2018})}\BibitemShut {NoStop}%
\bibitem [{\citenamefont {Koyuk}\ and\ \citenamefont
  {Seifert}(2019)}]{Operationally_Accessible}%
  \BibitemOpen
  \bibfield  {author} {\bibinfo {author} {\bibfnamefont {T.}~\bibnamefont
  {Koyuk}}\ and\ \bibinfo {author} {\bibfnamefont {U.}~\bibnamefont
  {Seifert}},\ }\bibfield  {title} {\bibinfo {title} {Operationally
  {A}ccessible {B}ounds on {F}luctuations and {E}ntropy {P}roduction in
  {P}eriodically {D}riven {S}ystems},\ }\href
  {https://doi.org/10.1103/PhysRevLett.122.230601} {\bibfield  {journal}
  {\bibinfo  {journal} {Phys. Rev. Lett.}\ }\textbf {\bibinfo {volume} {122}},\
  \bibinfo {pages} {230601} (\bibinfo {year} {2019})}\BibitemShut {NoStop}%
\bibitem [{\citenamefont {Saryal}\ \emph {et~al.}(2021)\citenamefont {Saryal},
  \citenamefont {Gerry}, \citenamefont {Khait}, \citenamefont {Segal},\ and\
  \citenamefont {Agarwalla}}]{Saryal_continuous}%
  \BibitemOpen
  \bibfield  {author} {\bibinfo {author} {\bibfnamefont {S.}~\bibnamefont
  {Saryal}}, \bibinfo {author} {\bibfnamefont {M.}~\bibnamefont {Gerry}},
  \bibinfo {author} {\bibfnamefont {I.}~\bibnamefont {Khait}}, \bibinfo
  {author} {\bibfnamefont {D.}~\bibnamefont {Segal}},\ and\ \bibinfo {author}
  {\bibfnamefont {B.~K.}\ \bibnamefont {Agarwalla}},\ }\bibfield  {title}
  {\bibinfo {title} {Universal {B}ounds on {F}luctuations in {C}ontinuous
  {T}hermal {M}achines},\ }\href
  {https://doi.org/10.1103/PhysRevLett.127.190603} {\bibfield  {journal}
  {\bibinfo  {journal} {Phys. Rev. Lett.}\ }\textbf {\bibinfo {volume} {127}},\
  \bibinfo {pages} {190603} (\bibinfo {year} {2021})}\BibitemShut {NoStop}%
\bibitem [{\citenamefont {Novikov}(1958)}]{novikov1958efficiency}%
  \BibitemOpen
  \bibfield  {author} {\bibinfo {author} {\bibfnamefont {I.~I.}\ \bibnamefont
  {Novikov}},\ }\bibfield  {title} {\bibinfo {title} {The efficiency of atomic
  power stations (a review)},\ }\href
  {https://www.sciencedirect.com/science/article/pii/0891391958902444}
  {\bibfield  {journal} {\bibinfo  {journal} {J. Nucl. Energy (1954)}\ }\textbf
  {\bibinfo {volume} {7}},\ \bibinfo {pages} {125} (\bibinfo {year}
  {1958})}\BibitemShut {NoStop}%
\bibitem [{\citenamefont {Curzon}\ and\ \citenamefont
  {Ahlborn}(1975)}]{curzon1975efficiency}%
  \BibitemOpen
  \bibfield  {author} {\bibinfo {author} {\bibfnamefont {F.~L.}\ \bibnamefont
  {Curzon}}\ and\ \bibinfo {author} {\bibfnamefont {B.}~\bibnamefont
  {Ahlborn}},\ }\bibfield  {title} {\bibinfo {title} {{E}fficiency of a
  {C}arnot engine at maximum power output},\ }\href
  {https://aapt.scitation.org/doi/abs/10.1119/1.10023} {\bibfield  {journal}
  {\bibinfo  {journal} {Am. J. Phys.}\ }\textbf {\bibinfo {volume} {43}},\
  \bibinfo {pages} {22} (\bibinfo {year} {1975})}\BibitemShut {NoStop}%
\bibitem [{\citenamefont {Hoffmann}\ \emph {et~al.}(1997)\citenamefont
  {Hoffmann}, \citenamefont {Burzler},\ and\ \citenamefont
  {Schubert}}]{Hoffman_review}%
  \BibitemOpen
  \bibfield  {author} {\bibinfo {author} {\bibfnamefont {K.}~\bibnamefont
  {Hoffmann}}, \bibinfo {author} {\bibfnamefont {J.}~\bibnamefont {Burzler}},\
  and\ \bibinfo {author} {\bibfnamefont {S.}~\bibnamefont {Schubert}},\
  }\bibfield  {title} {\bibinfo {title} {Endoreversible {T}hermodynamics},\
  }\href {https://doi.org/doi:10.1515/jnet.1997.22.4.311} {\bibfield  {journal}
  {\bibinfo  {journal} {J. Non-Equilib. Thermodyn.}\ }\textbf {\bibinfo
  {volume} {22}},\ \bibinfo {pages} {311} (\bibinfo {year} {1997})}\BibitemShut
  {NoStop}%
\bibitem [{\citenamefont {Andresen}(2011)}]{Andresen_review}%
  \BibitemOpen
  \bibfield  {author} {\bibinfo {author} {\bibfnamefont {B.}~\bibnamefont
  {Andresen}},\ }\bibfield  {title} {\bibinfo {title} {Current {T}rends in
  {F}inite-{T}ime {T}hermodynamics},\ }\href
  {https://doi.org/10.1002/anie.201001411} {\bibfield  {journal} {\bibinfo
  {journal} {Angew. Chem. Int. Ed.}\ }\textbf {\bibinfo {volume} {50}},\
  \bibinfo {pages} {2690} (\bibinfo {year} {2011})}\BibitemShut {NoStop}%
\bibitem [{\citenamefont {Bouton}\ \emph {et~al.}(2021)\citenamefont {Bouton},
  \citenamefont {Nettersheim}, \citenamefont {Burgardt}, \citenamefont {Adam},
  \citenamefont {Lutz},\ and\ \citenamefont {Widera}}]{bouton2021quantum}%
  \BibitemOpen
  \bibfield  {author} {\bibinfo {author} {\bibfnamefont {Q.}~\bibnamefont
  {Bouton}}, \bibinfo {author} {\bibfnamefont {J.}~\bibnamefont {Nettersheim}},
  \bibinfo {author} {\bibfnamefont {S.}~\bibnamefont {Burgardt}}, \bibinfo
  {author} {\bibfnamefont {D.}~\bibnamefont {Adam}}, \bibinfo {author}
  {\bibfnamefont {E.}~\bibnamefont {Lutz}},\ and\ \bibinfo {author}
  {\bibfnamefont {A.}~\bibnamefont {Widera}},\ }\bibfield  {title} {\bibinfo
  {title} {A quantum heat engine driven by atomic collisions},\ }\href
  {https://doi.org/10.1038/s41467-021-22222-z} {\bibfield  {journal} {\bibinfo
  {journal} {Nat. Commun.}\ }\textbf {\bibinfo {volume} {12}},\ \bibinfo
  {pages} {2063} (\bibinfo {year} {2021})}\BibitemShut {NoStop}%
\bibitem [{\citenamefont {Schmiedl}\ and\ \citenamefont
  {Seifert}(2007)}]{Schmiedl_2007}%
  \BibitemOpen
  \bibfield  {author} {\bibinfo {author} {\bibfnamefont {T.}~\bibnamefont
  {Schmiedl}}\ and\ \bibinfo {author} {\bibfnamefont {U.}~\bibnamefont
  {Seifert}},\ }\bibfield  {title} {\bibinfo {title} {Efficiency at maximum
  power: {A}n analytically solvable model for stochastic heat engines},\ }\href
  {https://doi.org/10.1209/0295-5075/81/20003} {\bibfield  {journal} {\bibinfo
  {journal} {EPL}\ }\textbf {\bibinfo {volume} {81}},\ \bibinfo {pages} {20003}
  (\bibinfo {year} {2007})}\BibitemShut {NoStop}%
\bibitem [{\citenamefont {Dechant}\ \emph {et~al.}(2015)\citenamefont
  {Dechant}, \citenamefont {Kiesel},\ and\ \citenamefont
  {Lutz}}]{DechantPRL2015}%
  \BibitemOpen
  \bibfield  {author} {\bibinfo {author} {\bibfnamefont {A.}~\bibnamefont
  {Dechant}}, \bibinfo {author} {\bibfnamefont {N.}~\bibnamefont {Kiesel}},\
  and\ \bibinfo {author} {\bibfnamefont {E.}~\bibnamefont {Lutz}},\ }\bibfield
  {title} {\bibinfo {title} {{A}ll-{O}ptical {N}anomechanical {H}eat
  {E}ngine},\ }\href {https://doi.org/10.1103/PhysRevLett.114.183602}
  {\bibfield  {journal} {\bibinfo  {journal} {Phys. Rev. Lett.}\ }\textbf
  {\bibinfo {volume} {114}},\ \bibinfo {pages} {183602} (\bibinfo {year}
  {2015})}\BibitemShut {NoStop}%
\bibitem [{\citenamefont {Dechant}\ \emph {et~al.}(2017)\citenamefont
  {Dechant}, \citenamefont {Kiesel},\ and\ \citenamefont
  {Lutz}}]{DechantEPL2017}%
  \BibitemOpen
  \bibfield  {author} {\bibinfo {author} {\bibfnamefont {A.}~\bibnamefont
  {Dechant}}, \bibinfo {author} {\bibfnamefont {N.}~\bibnamefont {Kiesel}},\
  and\ \bibinfo {author} {\bibfnamefont {E.}~\bibnamefont {Lutz}},\ }\bibfield
  {title} {\bibinfo {title} {Underdamped stochastic heat engine at maximum
  efficiency},\ }\href {https://doi.org/10.1209/0295-5075/119/50003} {\bibfield
   {journal} {\bibinfo  {journal} {EPL}\ }\textbf {\bibinfo {volume} {119}},\
  \bibinfo {pages} {50003} (\bibinfo {year} {2017})}\BibitemShut {NoStop}%
\bibitem [{\citenamefont {Deffner}(2018)}]{Deffner_entropy}%
  \BibitemOpen
  \bibfield  {author} {\bibinfo {author} {\bibfnamefont {S.}~\bibnamefont
  {Deffner}},\ }\bibfield  {title} {\bibinfo {title} {Efficiency of {H}armonic
  {Q}uantum {O}tto {E}ngines at {M}aximal {P}ower},\ }\href
  {https://doi.org/10.3390/e20110875} {\bibfield  {journal} {\bibinfo
  {journal} {Entropy}\ }\textbf {\bibinfo {volume} {20}},\ \bibinfo {pages}
  {875} (\bibinfo {year} {2018})}\BibitemShut {NoStop}%
\bibitem [{\citenamefont {Chen}\ \emph
  {et~al.}(2019{\natexlab{a}})\citenamefont {Chen}, \citenamefont {Sun},\ and\
  \citenamefont {Dong}}]{ChenJF_2}%
  \BibitemOpen
  \bibfield  {author} {\bibinfo {author} {\bibfnamefont {J.-F.}\ \bibnamefont
  {Chen}}, \bibinfo {author} {\bibfnamefont {C.-P.}\ \bibnamefont {Sun}},\ and\
  \bibinfo {author} {\bibfnamefont {H.}~\bibnamefont {Dong}},\ }\bibfield
  {title} {\bibinfo {title} {Achieve higher efficiency at maximum power with
  finite-time quantum {O}tto cycle},\ }\href
  {https://doi.org/10.1103/PhysRevE.100.062140} {\bibfield  {journal} {\bibinfo
   {journal} {Phys. Rev. E}\ }\textbf {\bibinfo {volume} {100}},\ \bibinfo
  {pages} {062140} (\bibinfo {year} {2019}{\natexlab{a}})}\BibitemShut
  {NoStop}%
\bibitem [{\citenamefont {Chen}\ \emph
  {et~al.}(2019{\natexlab{b}})\citenamefont {Chen}, \citenamefont {Sun},\ and\
  \citenamefont {Dong}}]{ChenJF_1}%
  \BibitemOpen
  \bibfield  {author} {\bibinfo {author} {\bibfnamefont {J.-F.}\ \bibnamefont
  {Chen}}, \bibinfo {author} {\bibfnamefont {C.-P.}\ \bibnamefont {Sun}},\ and\
  \bibinfo {author} {\bibfnamefont {H.}~\bibnamefont {Dong}},\ }\bibfield
  {title} {\bibinfo {title} {Boosting the performance of quantum {O}tto heat
  engines},\ }\href {https://doi.org/10.1103/PhysRevE.100.032144} {\bibfield
  {journal} {\bibinfo  {journal} {Phys. Rev. E}\ }\textbf {\bibinfo {volume}
  {100}},\ \bibinfo {pages} {032144} (\bibinfo {year}
  {2019}{\natexlab{b}})}\BibitemShut {NoStop}%
\bibitem [{\citenamefont {Izumida}(2022)}]{Izumida_2204.00807}%
  \BibitemOpen
  \bibfield  {author} {\bibinfo {author} {\bibfnamefont {Y.}~\bibnamefont
  {Izumida}},\ }\bibfield  {title} {\bibinfo {title} {{Irreversible efficiency
  and Carnot theorem for heat engines operating with multiple heat baths in
  linear response regime}},\ }\href
  {https://doi.org/10.1103/PhysRevResearch.4.023217} {\bibfield  {journal}
  {\bibinfo  {journal} {Phys. Rev. Research}\ }\textbf {\bibinfo {volume}
  {4}},\ \bibinfo {pages} {023217} (\bibinfo {year} {2022})}\BibitemShut
  {NoStop}%
\bibitem [{\citenamefont {Landsberg}\ and\ \citenamefont
  {Leff}(1989)}]{Landsberg}%
  \BibitemOpen
  \bibfield  {author} {\bibinfo {author} {\bibfnamefont {P.~T.}\ \bibnamefont
  {Landsberg}}\ and\ \bibinfo {author} {\bibfnamefont {H.~S.}\ \bibnamefont
  {Leff}},\ }\bibfield  {title} {\bibinfo {title} {Thermodynamic cycles with
  nearly universal maximum-work efficiencies},\ }\href
  {https://doi.org/10.1088/0305-4470/22/18/034} {\bibfield  {journal} {\bibinfo
   {journal} {J. Phys. A: Math. Gen.}\ }\textbf {\bibinfo {volume} {22}},\
  \bibinfo {pages} {4019} (\bibinfo {year} {1989})}\BibitemShut {NoStop}%
\bibitem [{Note1()}]{Note1}%
  \BibitemOpen
  \bibinfo {note} {The term ``reversible'' in this work means that the working
  substance of the engine is reversible such that the working substance is
  always in a canonical state for some temperature. Following the assumption of
  endoreversible thermodynamics, we regard that the working substance is
  reversible, but the heat flow between the working substance and the bath can
  be irreversible.}\BibitemShut {Stop}%
\bibitem [{Note2()}]{Note2}%
  \BibitemOpen
  \bibinfo {note} {There are many cases that the heat capacities of the heat
  transfer strokes are constant. For example, isobaric and isochoric strokes
  for a Brownian particle in a harmonic oscillator potential or a square-well
  potential.}\BibitemShut {Stop}%
\bibitem [{\citenamefont {Sato}\ \emph {et~al.}(2002)\citenamefont {Sato},
  \citenamefont {Sekimoto}, \citenamefont {Hondou},\ and\ \citenamefont
  {Takagi}}]{sato_rev_cond}%
  \BibitemOpen
  \bibfield  {author} {\bibinfo {author} {\bibfnamefont {K.}~\bibnamefont
  {Sato}}, \bibinfo {author} {\bibfnamefont {K.}~\bibnamefont {Sekimoto}},
  \bibinfo {author} {\bibfnamefont {T.}~\bibnamefont {Hondou}},\ and\ \bibinfo
  {author} {\bibfnamefont {F.}~\bibnamefont {Takagi}},\ }\bibfield  {title}
  {\bibinfo {title} {Irreversibility resulting from contact with a heat bath
  caused by the finiteness of the system},\ }\href
  {https://doi.org/10.1103/PhysRevE.66.016119} {\bibfield  {journal} {\bibinfo
  {journal} {Phys. Rev. E}\ }\textbf {\bibinfo {volume} {66}},\ \bibinfo
  {pages} {016119} (\bibinfo {year} {2002})}\BibitemShut {NoStop}%
\bibitem [{\citenamefont {Qian}(2001)}]{Qian}%
  \BibitemOpen
  \bibfield  {author} {\bibinfo {author} {\bibfnamefont {H.}~\bibnamefont
  {Qian}},\ }\bibfield  {title} {\bibinfo {title} {Mesoscopic nonequilibrium
  thermodynamics of single macromolecules and dynamic entropy-energy
  compensation},\ }\href {https://doi.org/10.1103/PhysRevE.65.016102}
  {\bibfield  {journal} {\bibinfo  {journal} {Phys. Rev. E}\ }\textbf {\bibinfo
  {volume} {65}},\ \bibinfo {pages} {016102} (\bibinfo {year}
  {2001})}\BibitemShut {NoStop}%
\bibitem [{Note3()}]{Note3}%
  \BibitemOpen
  \bibinfo {note} {The Hamiltonian in Ref.~\cite {Holubec_CTUR} is
  $H=kx^{2n}/2n$, where $x$ is the position of the Brownian particle, $k$ is
  the stiffness of the potential, and $n$ is a natural number. In this case,
  $C_V$ is constant, $C_V = k_B/(2n)$, which is in accordance with the
  assumption of our analysis. Therefore, Eq.~\protect \textup {\hbox
  {\mathsurround \z@ \protect \normalfont (\ignorespaces \ref
  {work_fluct}\unskip \@@italiccorr )}} is applicable to this case, and gives
  $\protect \sqrt {\sigma _W} = k_B/(\protect \sqrt {n}\protect \tmspace
  +\thinmuskip {.1667em}\Delta S)$, which agrees with the result in Ref.~\cite
  {Holubec_CTUR}.}\BibitemShut {Stop}%
\bibitem [{\citenamefont {Gardiner}(2004)}]{handbook}%
  \BibitemOpen
  \bibfield  {author} {\bibinfo {author} {\bibfnamefont {C.~W.}\ \bibnamefont
  {Gardiner}},\ }\href@noop {} {\emph {\bibinfo {title} {Handbook of Stochastic
  Methods}}},\ \bibinfo {edition} {3rd}\ ed.\ (\bibinfo  {publisher} {Springer,
  Berlin},\ \bibinfo {year} {2004})\BibitemShut {NoStop}%
\bibitem [{\citenamefont {Xu}\ and\ \citenamefont
  {Watanabe}(2022)}]{xu2021correlationenhanced}%
  \BibitemOpen
  \bibfield  {author} {\bibinfo {author} {\bibfnamefont {G.-H.}\ \bibnamefont
  {Xu}}\ and\ \bibinfo {author} {\bibfnamefont {G.}~\bibnamefont {Watanabe}},\
  }\bibfield  {title} {\bibinfo {title} {Correlation-enhanced stability of
  microscopic cyclic heat engines},\ }\href
  {https://doi.org/10.1103/PhysRevResearch.4.L032017} {\bibfield  {journal}
  {\bibinfo  {journal} {Phys. Rev. Research}\ }\textbf {\bibinfo {volume}
  {4}},\ \bibinfo {pages} {L032017} (\bibinfo {year} {2022})}\BibitemShut
  {NoStop}%
\bibitem [{\citenamefont {Imparato}\ \emph {et~al.}(2007)\citenamefont
  {Imparato}, \citenamefont {Peliti}, \citenamefont {Pesce}, \citenamefont
  {Rusciano},\ and\ \citenamefont {Sasso}}]{PDF_1}%
  \BibitemOpen
  \bibfield  {author} {\bibinfo {author} {\bibfnamefont {A.}~\bibnamefont
  {Imparato}}, \bibinfo {author} {\bibfnamefont {L.}~\bibnamefont {Peliti}},
  \bibinfo {author} {\bibfnamefont {G.}~\bibnamefont {Pesce}}, \bibinfo
  {author} {\bibfnamefont {G.}~\bibnamefont {Rusciano}},\ and\ \bibinfo
  {author} {\bibfnamefont {A.}~\bibnamefont {Sasso}},\ }\bibfield  {title}
  {\bibinfo {title} {Work and heat probability distribution of an optically
  driven {B}rownian particle: {T}heory and experiments},\ }\href
  {https://doi.org/10.1103/PhysRevE.76.050101} {\bibfield  {journal} {\bibinfo
  {journal} {Phys. Rev. E}\ }\textbf {\bibinfo {volume} {76}},\ \bibinfo
  {pages} {050101(R)} (\bibinfo {year} {2007})}\BibitemShut {NoStop}%
\bibitem [{\citenamefont {Chatterjee}\ and\ \citenamefont
  {Cherayil}(2010)}]{PDF_2}%
  \BibitemOpen
  \bibfield  {author} {\bibinfo {author} {\bibfnamefont {D.}~\bibnamefont
  {Chatterjee}}\ and\ \bibinfo {author} {\bibfnamefont {B.~J.}\ \bibnamefont
  {Cherayil}},\ }\bibfield  {title} {\bibinfo {title} {Exact path-integral
  evaluation of the heat distribution function of a trapped {B}rownian
  oscillator},\ }\href {https://doi.org/10.1103/PhysRevE.82.051104} {\bibfield
  {journal} {\bibinfo  {journal} {Phys. Rev. E}\ }\textbf {\bibinfo {volume}
  {82}},\ \bibinfo {pages} {051104} (\bibinfo {year} {2010})}\BibitemShut
  {NoStop}%
\bibitem [{\citenamefont {Gomez-Solano}\ \emph {et~al.}(2011)\citenamefont
  {Gomez-Solano}, \citenamefont {Petrosyan},\ and\ \citenamefont
  {Ciliberto}}]{PDF_exp}%
  \BibitemOpen
  \bibfield  {author} {\bibinfo {author} {\bibfnamefont {J.~R.}\ \bibnamefont
  {Gomez-Solano}}, \bibinfo {author} {\bibfnamefont {A.}~\bibnamefont
  {Petrosyan}},\ and\ \bibinfo {author} {\bibfnamefont {S.}~\bibnamefont
  {Ciliberto}},\ }\bibfield  {title} {\bibinfo {title} {Heat {F}luctuations in
  a {N}onequilibrium {B}ath},\ }\href
  {https://doi.org/10.1103/PhysRevLett.106.200602} {\bibfield  {journal}
  {\bibinfo  {journal} {Phys. Rev. Lett.}\ }\textbf {\bibinfo {volume} {106}},\
  \bibinfo {pages} {200602} (\bibinfo {year} {2011})}\BibitemShut {NoStop}%
\bibitem [{\citenamefont {Pal}\ and\ \citenamefont
  {Sabhapandit}(2013)}]{PDF_3}%
  \BibitemOpen
  \bibfield  {author} {\bibinfo {author} {\bibfnamefont {A.}~\bibnamefont
  {Pal}}\ and\ \bibinfo {author} {\bibfnamefont {S.}~\bibnamefont
  {Sabhapandit}},\ }\bibfield  {title} {\bibinfo {title} {Work fluctuations for
  a {B}rownian particle in a harmonic trap with fluctuating locations},\ }\href
  {https://doi.org/10.1103/PhysRevE.87.022138} {\bibfield  {journal} {\bibinfo
  {journal} {Phys. Rev. E}\ }\textbf {\bibinfo {volume} {87}},\ \bibinfo
  {pages} {022138} (\bibinfo {year} {2013})}\BibitemShut {NoStop}%
\bibitem [{\citenamefont {Salazar}\ and\ \citenamefont {Lira}(2016)}]{PDF_4}%
  \BibitemOpen
  \bibfield  {author} {\bibinfo {author} {\bibfnamefont {D.~S.~P.}\
  \bibnamefont {Salazar}}\ and\ \bibinfo {author} {\bibfnamefont {S.~A.}\
  \bibnamefont {Lira}},\ }\bibfield  {title} {\bibinfo {title} {Exactly
  solvable nonequilibrium {L}angevin relaxation of a trapped nanoparticle},\
  }\href {https://doi.org/10.1088/1751-8113/49/46/465001} {\bibfield  {journal}
  {\bibinfo  {journal} {J. Phys. A: Math. Theor.}\ }\textbf {\bibinfo {volume}
  {49}},\ \bibinfo {pages} {465001} (\bibinfo {year} {2016})}\BibitemShut
  {NoStop}%
\bibitem [{\citenamefont {Crisanti}\ \emph {et~al.}(2017)\citenamefont
  {Crisanti}, \citenamefont {Sarracino},\ and\ \citenamefont
  {Zannetti}}]{PDF_5}%
  \BibitemOpen
  \bibfield  {author} {\bibinfo {author} {\bibfnamefont {A.}~\bibnamefont
  {Crisanti}}, \bibinfo {author} {\bibfnamefont {A.}~\bibnamefont
  {Sarracino}},\ and\ \bibinfo {author} {\bibfnamefont {M.}~\bibnamefont
  {Zannetti}},\ }\bibfield  {title} {\bibinfo {title} {Heat fluctuations of
  {B}rownian oscillators in nonstationary processes: {F}luctuation theorem and
  condensation transition},\ }\href
  {https://doi.org/10.1103/PhysRevE.95.052138} {\bibfield  {journal} {\bibinfo
  {journal} {Phys. Rev. E}\ }\textbf {\bibinfo {volume} {95}},\ \bibinfo
  {pages} {052138} (\bibinfo {year} {2017})}\BibitemShut {NoStop}%
\bibitem [{\citenamefont {Goswami}(2019)}]{PDF_6}%
  \BibitemOpen
  \bibfield  {author} {\bibinfo {author} {\bibfnamefont {K.}~\bibnamefont
  {Goswami}},\ }\bibfield  {title} {\bibinfo {title} {Heat fluctuation of a
  harmonically trapped particle in an active bath},\ }\href
  {https://doi.org/10.1103/PhysRevE.99.012112} {\bibfield  {journal} {\bibinfo
  {journal} {Phys. Rev. E}\ }\textbf {\bibinfo {volume} {99}},\ \bibinfo
  {pages} {012112} (\bibinfo {year} {2019})}\BibitemShut {NoStop}%
\bibitem [{\citenamefont {Chvosta}\ \emph {et~al.}(2020)\citenamefont
  {Chvosta}, \citenamefont {Lips}, \citenamefont {Holubec}, \citenamefont
  {Ryabov},\ and\ \citenamefont {Maass}}]{PDF_7}%
  \BibitemOpen
  \bibfield  {author} {\bibinfo {author} {\bibfnamefont {P.}~\bibnamefont
  {Chvosta}}, \bibinfo {author} {\bibfnamefont {D.}~\bibnamefont {Lips}},
  \bibinfo {author} {\bibfnamefont {V.}~\bibnamefont {Holubec}}, \bibinfo
  {author} {\bibfnamefont {A.}~\bibnamefont {Ryabov}},\ and\ \bibinfo {author}
  {\bibfnamefont {P.}~\bibnamefont {Maass}},\ }\bibfield  {title} {\bibinfo
  {title} {Statistics of work performed by optical tweezers with general
  time-variation of their stiffness},\ }\href
  {https://doi.org/10.1088/1751-8121/ab95c2} {\bibfield  {journal} {\bibinfo
  {journal} {J. Phys. A: Math. Theor.}\ }\textbf {\bibinfo {volume} {53}},\
  \bibinfo {pages} {275001} (\bibinfo {year} {2020})}\BibitemShut {NoStop}%
\bibitem [{\citenamefont {Crooks}(2007)}]{Crook_geo}%
  \BibitemOpen
  \bibfield  {author} {\bibinfo {author} {\bibfnamefont {G.~E.}\ \bibnamefont
  {Crooks}},\ }\bibfield  {title} {\bibinfo {title} {Measuring {T}hermodynamic
  {L}ength},\ }\href {https://doi.org/10.1103/PhysRevLett.99.100602} {\bibfield
   {journal} {\bibinfo  {journal} {Phys. Rev. Lett.}\ }\textbf {\bibinfo
  {volume} {99}},\ \bibinfo {pages} {100602} (\bibinfo {year}
  {2007})}\BibitemShut {NoStop}%
\bibitem [{\citenamefont {Van~Vu}\ and\ \citenamefont
  {Hasegawa}(2021)}]{Van_geo}%
  \BibitemOpen
  \bibfield  {author} {\bibinfo {author} {\bibfnamefont {T.}~\bibnamefont
  {Van~Vu}}\ and\ \bibinfo {author} {\bibfnamefont {Y.}~\bibnamefont
  {Hasegawa}},\ }\bibfield  {title} {\bibinfo {title} {Geometrical bounds of
  the irreversibility in markovian systems},\ }\href
  {https://doi.org/10.1103/PhysRevLett.126.010601} {\bibfield  {journal}
  {\bibinfo  {journal} {Phys. Rev. Lett.}\ }\textbf {\bibinfo {volume} {126}},\
  \bibinfo {pages} {010601} (\bibinfo {year} {2021})}\BibitemShut {NoStop}%
\bibitem [{\citenamefont {Brandner}\ and\ \citenamefont
  {Saito}(2020)}]{Brandner_geo}%
  \BibitemOpen
  \bibfield  {author} {\bibinfo {author} {\bibfnamefont {K.}~\bibnamefont
  {Brandner}}\ and\ \bibinfo {author} {\bibfnamefont {K.}~\bibnamefont
  {Saito}},\ }\bibfield  {title} {\bibinfo {title} {Thermodynamic {G}eometry of
  {M}icroscopic {H}eat {E}ngines},\ }\href
  {https://doi.org/10.1103/PhysRevLett.124.040602} {\bibfield  {journal}
  {\bibinfo  {journal} {Phys. Rev. Lett.}\ }\textbf {\bibinfo {volume} {124}},\
  \bibinfo {pages} {040602} (\bibinfo {year} {2020})}\BibitemShut {NoStop}%
\bibitem [{\citenamefont {Frim}\ and\ \citenamefont
  {DeWeese}(2022)}]{Frim_geo}%
  \BibitemOpen
  \bibfield  {author} {\bibinfo {author} {\bibfnamefont {A.~G.}\ \bibnamefont
  {Frim}}\ and\ \bibinfo {author} {\bibfnamefont {M.~R.}\ \bibnamefont
  {DeWeese}},\ }\bibfield  {title} {\bibinfo {title} {{Optimal finite-time
  Brownian Carnot engine}},\ }\href
  {https://doi.org/10.1103/PhysRevE.105.L052103} {\bibfield  {journal}
  {\bibinfo  {journal} {Phys. Rev. E}\ }\textbf {\bibinfo {volume} {105}},\
  \bibinfo {pages} {L052103} (\bibinfo {year} {2022})}\BibitemShut {NoStop}%
\end{thebibliography}%

\end{document}